\documentclass[aps,prd,10pt,nofootinbib,onecolumn,a4paper]{revtex4-2}

\usepackage{bm,latexsym,amsmath,amssymb,amsfonts,mathrsfs,amsfonts,mathtools}
\usepackage{graphicx}
\usepackage{xcolor}
\usepackage[hang,small,bf]{caption}
\usepackage[subrefformat=parens]{subcaption}
\usepackage{ascmac}
\usepackage{physics}
\usepackage{color}
\usepackage{float}
\usepackage{braket}
\usepackage{bbm}
\usepackage{hyperref}
\usepackage{comment} 

\newcommand{\simgt}{\lower.5ex\hbox{$\; \buildrel > \over \sim \;$}}
\newcommand{\simlt}{\lower.5ex\hbox{$\; \buildrel < \over \sim \;$}}

\begin{document}

\title{Testing classical–quantum gravity with geodesic deviation}

\author{Tomoya Hirotani}
\email{hirotani.tomoya.937@s.kyushu-u.ac.jp}
\affiliation{Department of Physics,  Kyushu University, 744 Motooka, Nishi-Ku, Fukuoka 819-0395, Japan}
\author{Akira Matsumura}
\email{matsumura.akira@phys.kyushu-u.ac.jp}
\affiliation{Department of Physics,  Kyushu University, 744 Motooka, Nishi-Ku, Fukuoka 819-0395, Japan}
\affiliation{Quantum and Spacetime Research Institute, Kyushu University, 744 Motooka, Nishi-Ku, Fukuoka 819-0395, Japan}

\begin{abstract}
A novel semiclassical gravity model proposed by Oppenheim \emph{et al}., that consistently describes interactions between quantum systems and a classical gravitational field, has recently attracted considerable attention. However, the limitations and phenomenological viability of this model have not yet been thoroughly investigated.
In this work, based on the model, we study quantum fluctuations of geodesic deviation coupled with a classical gravitational field. We analytically derive the strain spectrum expected from the fluctuations and show that the original Oppenheim \emph{et al}. model can be tested with the current observational sensitivity of gravitational-wave experiments.
Furthermore, motivated by the novel semiclassical model, we construct two additional models: a modified Oppenheim \emph{et al}. model that is manifestly consistent with Bianchi identities, and a classical-quantum model with environment-induced noise.
We analyze the strain spectra predicted by these two models through comparison with those of the original Oppenheim \emph{et al.} model and perturbative quantum gravity.
\end{abstract}

\maketitle

\section{Introduction}

Probing the quantum nature of gravity is one of the most important goals in modern physics. 
Many previous studies have theoretically predicted characteristic phenomena that would arise if gravity were quantum mechanical, 
and have aimed to test these predictions using future or ongoing experiments. 
However, despite decades of theoretical and experimental efforts, no definitive conclusion has yet been reached as to whether gravity is fundamentally quantum in nature.

Against this backdrop, renewed attention has recently been paid to theoretical frameworks based on the standpoint that gravity is fundamentally classical, 
while quantum properties are attributed solely to matter fields. 
In fact this line of thought is not new: already in the 1960s, Møller and Rosenfeld proposed the semiclassical Einstein equation, 
in which the expectation value of the stress--energy tensor of quantum matter serves as the source of a classical gravitational field~\cite{moller1962theories,rosenfeld1963quantization},
\begin{align}
    G_{\mu\nu}(x) = 8\pi G_N \braket{\hat T_{\mu\nu}(x)}.
    \label{semiclassical_Einstein}
\end{align}
As one of the earliest attempts to self-consistently couple quantum fields to a classical gravitational field, 
this framework laid the foundation of semiclassical gravity.

Subsequently, Schr\"odinger--Newton (SN) equation emerged as a concrete realization of this semiclassical coupling in the nonrelativistic limit. 
The SN equation gives the Newtonian gravitational potential depending on the state of quantum matter, which leads to a nonlinear Schr\"odinger equation.
This form originally appeared naturally in the study of self-gravitating Bose condensates (boson stars) by Ruffini and Bonazzola~\cite{PhysRev.187.1767}. 
However, the SN model suffers from a fundamental problem, for example, the causality violation originated from the nonlinearity of the Schr\"{o}dinger equation \cite{Diosi_2016,GISIN1990}. 
Furthermore, since both the semiclassical Einstein equation given by Eq.~\eqref{semiclassical_Einstein} and the SN model describe gravity as a deterministic force, these models are not valid when the energy-momentum tensor has large quantum fluctuations. 
By focusing on this feature, the limitation of the SN model was experimentally tested in Ref.~\cite{PageGeilker1981}.

To incorporate the effects of stress--energy fluctuations beyond the mean-field description, the framework of stochastic gravity was developed, in which the semiclassical Einstein equation is extended by introducing a stochastic source term characterized by the noise kernel of quantum matter fields~\cite{PhysRevD.47.4510,Hu2008_Stochastic_Gravity,PhysRevD.70.044002,PhysRevD.61.124024,PhysRevD.49.6636,Verdaguer2006StochasticGB,Guillem_Perez-Nadal_2010,PhysRevD.56.2163,PhysRevD.105.086004,Ford2022}. In this approach, the fluctuations of the stress--energy tensor induce stochastic fluctuations of the spacetime metric, providing a systematic extension of semiclassical gravity beyond the deterministic mean-field description.
In addition, the validity of these semiclassical gravity models is discussed in Ref.~\cite{PhysRevD.70.044002} from the viewpoint of the stability of the solutions with respect to metric fluctuations.

Later, semiclassical gravity models in the nonrelativistic Newtonian regime were independently reintroduced by Di\'osi and Penrose 
as models to discuss gravity-induced 
wavefunction collapse~\cite{DIOSI1984199,Penrose1996}. 
By endowing the gravitational field with stochasticity as above, these DP models avoid the violation of causality; 
nevertheless, they too have increasingly been challenged by experimental result \cite{Donadi2021}. 
Subsequently, Kafri, Taylor, and Milburn proposed a semiclassical gravity model based on an information-theoretic perspective, in which gravitational interaction is treated as an information channel consisting of classical measurement and feedback processes \cite{Kafri_2014}. This framework, however, has also been subjected to strong experimental constraints \cite{KFMTD_experiment}.

Against this background, a model recently proposed by Oppenheim \emph{et al.}~\cite{PhysRevX.13.041040,oppenheim2025covariant,grudka2024Renormalisation,Oppenheim2023_Nature,Layton2024healthiersemi} provides a new theoretical framework for consistently coupling quantum matter to gravity while keeping spacetime fundamentally classical. 
A central feature of this model is the existence of a trade-off relation between the magnitude of gravity-induced quantum decoherence and the diffusion associated with stochastic fluctuations of the gravitational field. 
Because this trade-off relation can lead to potentially observable effects even in low-energy and non-relativistic regimes, the model has attracted significant attention as a theoretical guideline for indirectly probing the quantum nature of gravity.
However, the properties and limitations of this model have not yet been fully understood.

To understand them concretely, in this work, we investigate the fluctuations of geodesic deviation between quantum objects in a classical gravitational field within the model proposed by Oppenheim \emph{et al}. 
There have been many studies investigating the effects of fluctuations in quantum gravitational fields on interferometers. 
For example, Refs.~\cite{PhysRevD.103.044017,PhysRevD.104.046021} examined the spectrum of fluctuations in the geodesic deviation induced by gravitons, while Refs.~\cite{Carney2024response,Freidel2026geometric} derived the spectrum of time-delay fluctuations in interferometers caused by gravitons. 
In the present paper, we analytically derive the strain spectra estimated from the fluctuations of deviation and clarify its physical characteristics.

In the original model by Oppenheim \emph{et al}., the classical gravitational field is excited by a stochastic source characterized by a noise kernel. However, we find that this noise kernel may not satisfy the Bianchi identities. We therefore propose a modified version of the Oppenheim \emph{et al}. model consistent with the Bianchi identities.
Furthermore, we construct another new model that incorporates environment-induced noise, which provides a possible origin of the stochastic source.
This model is phenomenologically introduced for capturing a situation in which a quantum gravitational field interacts with an environmental quantum field, leading to effective fluctuations of the gravitational field.
Based on the proposed new models, 
we perform the same spectral analysis 
and examine their behavior through  comparison with those predicted by the original Oppenheim \emph{et al}. model and perturbative quantum gravity. 
We also discuss the constraints of the model by
Oppenheim \emph{et al}. and our proposed ones from current and future experiments involving gravitational-wave detectors.

The structure of this paper is as follows. In Sec.~\ref{sec:Classical_Quantum_dynamics}, we review the formulation of classical–quantum dynamics proposed by Oppenheim \emph{et al}. in Refs.~\cite{PhysRevX.13.041040,oppenheim2025covariant,grudka2024Renormalisation,Oppenheim2023_Nature}. In Sec.~\ref{sec:Geodesic_deviation}, in order to obtain the strain spectrum of the geodesic deviation, we first introduce the action for the geodesic deviation. We then derive the dynamics of the total system in the path-integral representation using the method outlined in Sec.~\ref{sec:Classical_Quantum_dynamics}. By applying the Feynman–Vernon influence functional approach and focusing on the geodesic deviation, we finally derive the corresponding Langevin equation.
In Sec.~\ref{sec:Noise_spectra}, we derive the strain spectra and discuss its characteristic features. In Sec.~\ref{sec:Experimental_constraints}, using the obtained spectra, we examine the expected constraints on the models considered in the present paper by current and future experiments.
Throughout this paper, we adopt natural units with $c=\hbar=1$, and use the mostly-plus convention for the Minkowski metric, $\eta_{\mu\nu}=\mathrm{diag}(-1,1,1,1)$.
Also, raising and lowering spacetime indices are defined by the Minkowski metric.

\section{Classical-Quantum dynamics}
\label{sec:Classical_Quantum_dynamics}
The problem of how to consistently describe the time evolution of a system in which classical and quantum degrees of freedom interact has long been discussed. In particular, to coherently combine the frameworks of classical probability theory and quantum mechanics, several approaches have been proposed that treat both stochastic classical variables and density operators dynamically. In Refs.~\cite{PhysRevX.13.041040,oppenheim2022classeshybridclassicalquantumdynamics}, the dynamics of such a classical-quantum(CQ) coupled system is formulated in the form of a completely positive and trace-preserving (CPTP) master equation, ensuring the consistent time evolution of the classical probability distribution and the quantum state. In this section, we briefly review the method proposed by Oppenheim \emph{et al}. in Ref.~\cite{oppenheim2025covariant} for describing the dynamics of CQ systems. 

A quantum state is generally represented by a density operator $\hat\rho(t)$, while a classical stochastic system is represented by a probability distribution $P(z, t)$, where $z$ denotes the set of classical degrees of freedom. We then define the CQ state generally as
\begin{equation}
    \hat\varrho(z,t) = \hat\rho(z,t) P(z,t),
    \label{CQ_state}
\end{equation}
which encodes statistical correlations mediated by the classical variable $z$.
This CQ state yields a normalized classical probability distribution
\begin{align}
    \operatorname{Tr}_{\mathcal{H}}[\hat\varrho(z,t)] = P(z,t),\notag
\end{align}
when taking the trace over the quantum variables, and a normalized density operator
\begin{align}
    \int dz \hat\varrho(z,t) = \hat\rho(t),\notag
\end{align}
when summing over the classical variables.

For the dynamics of this system, in order to make it applicable to field-theoretic formulations and to impose symmetries more conveniently, we use a path integral formalism.
The quantum subsystem evolves via the Schwinger–Keldysh path integral \cite{Milton2015SchwingerAction}, while the classical subsystem evolves according to the Fokker–Planck path integral \cite{oppenheim2025covariant}. To be consistent with such path integrals, the CQ state evolves through the path integral
\begin{align}
    \varrho[q_f, \underline{q_f}, z_f, t_f] = \int \mathcal{D}q \mathcal{D}\underline{q} \mathcal{D}z e^{I_{\mathrm{CQ}}[q,\underline{q},z]} \varrho[q_i, \underline{q_i}, z_i, t_i],
    \label{CQ_full_dynamics}
\end{align}
where $\varrho[q, \underline{q}, z, t] = \bra{q} \hat\varrho(z,t) \ket{\underline{q}}$ represents the component of the CQ state \eqref{CQ_state}, expressed using the forward branch $q$ and the backward branch $\underline{q}$ of the quantum system. Here, the total action of the system denoted by $I_{\mathrm{CQ}}$ is 
\begin{align}
    I_{\mathrm{CQ}}[q,\underline{q},z] = iS_\text{tot}[q,z] - iS_\text{tot}[\underline{q},z] -\int dt dt' \sum_{i,j} \left[
     \frac{1}{2}\frac{\delta \Delta S_{\mathrm{CQ}}}{\delta z_i(t)} D_{ij}(t,t') \frac{\delta \Delta S_{\mathrm{CQ}}}{\delta z_j(t')} +\frac{1}{2}\frac{\delta \bar{S}_{\mathrm{CQ}}}{\delta z_i(t)} N_{ij}^{-1}(t,t') \frac{\delta \bar{S}_{\mathrm{CQ}}}{\delta z_j(t')}
    \right ],
    \label{general_CQ_action}
\end{align}
where we defined $\Delta S_{\mathrm{CQ}} \equiv S_{\mathrm{tot}}[q,z] - S_{\mathrm{tot}}[\underline{q},z]$, $\bar{S}_{\mathrm{CQ}} \equiv \frac{1}{2}\left( S_{\mathrm{tot}}[q,z] + S_{\mathrm{tot}}[\underline{q},z] \right)$ from the action of the total system  $S_{tot}$.
The functions $D_{ij} (t,t')$ and $N_{ij}(t,t')$ are symmetric and positive semi-definite matrices. 
In Eq.~\eqref{general_CQ_action}, the terms involving $D_{ij}(t,t')$ are related to the decoherence of the quantum system, 
whereas the terms involving $N_{ij}(t,t')$ are associated with diffusion in the classical system.
In order to preserve the completely positivity in the path integral Eq.~\eqref{CQ_full_dynamics}, there exists a \textit{decoherence-diffusion trade-off relation} between $D_{ij}(t,t')$ and $N_{ij}(t,t')$  given by
\begin{align}
    D N \ge \frac{1}{4},
    \label{trade_off}
\end{align}
where this is a matrix inequality for the matrices $D_{ij}(t,t')$ and $N_{ij}(t,t')$. 
The inequality reflects the following property: when the effect of noise in the classical system is small, 
the quantum system undergoes strong decoherence; conversely, for the decoherence of the quantum system 
to be suppressed, the diffusion in the classical system must be sufficiently large.

\section{Geodesic deviation}
\label{sec:Geodesic_deviation}

In this section, we present our model.
We assume a point mass $M$ located at the spacetime point $x^\mu$ and a point mass $m$ located at $y^\mu$, which are coupled to a gravitational field.
The total action of the masses and the gravitational field is written as
\begin{align}
    S_{\mathrm{tot}}=-M\int d\lambda \sqrt{-g_{\mu\nu}(x)\frac{dx^{\mu}}{d\lambda} \frac{dx^{\nu}}{d\lambda}}
    -
    m\int d\lambda \sqrt{-g_{\mu\nu}(y)\frac{dy^{\mu}}{d\lambda} \frac{dy^{\nu}}{d\lambda}}
    + \frac{1}{16\pi G_N} \int d^4x \, \sqrt{-g}\, R.
    \label{total_action_geoEH}
\end{align}
We consider the small deviation $\xi^\mu$ between the masses,
\begin{equation}
    y^{\mu} = x^{\mu} + \xi^{\mu},\notag
\end{equation}
and the metric perturbations around the flat spacetime $g_{\mu\nu} = \eta_{\mu\nu} + h_{\mu\nu}$. 
Mass $M$ is assumed to be almost at rest,  $x^\mu \sim X^\mu(t)=(t,0,0,0)$.  
Expanding the action with respect to $\xi^{\mu}$ and $h_{\mu\nu}$ up to second order and writing the action in the Fermi normal coordinates~\cite{PhysRevD.103.044017}, we obtain the effective action of the deviation and the metric perturbations as 
\begin{align}
    S_\text{tot}=\frac{m}{2}\int dt \left( \delta_{ab} \frac{d\xi^{a}}{dt}\frac{d\xi^{b}}{dt}
    -
    R^{(1)}_{0a0b} \xi^a \xi^b \right)  + \frac{1}{32\pi G_N} \int d^4x \, \Big[-\frac{1}{2}\partial_\alpha h_{\mu\nu} \partial^\alpha h^{\mu\nu}+\frac{1}{2} \partial_\alpha h \partial^\alpha h- \partial_\alpha h \partial_\beta h^{\alpha \beta}+\partial_\alpha h^{\alpha\mu} \partial_\beta h^{\beta}_{\mu}\Big],
    \label{effective_action}
\end{align}
where $h=\eta^{\mu\nu} h_{\mu\nu}$, $\delta_{ab} \,(a,b=x,y,z)$ is the Kronecker delta, and $\xi^{a}$ is the perpendicular component of $\xi^\mu$ to mass $M$ four-velocity $U^{\mu}=dX^\mu/dt=(1,0,0,0)$. 
The tensor
$R^{(1)}_{0a0b}$ is the Riemann curvature taken up to first order in $h_{\mu\nu}$.
The first part of the action, $S_\text{dev}=\frac{m}{2}\int dt \left( \delta_{ab} \frac{d\xi^{a}}{dt}\frac{d\xi^{b}}{dt}-R^{(1)}_{0a0b} \xi^a \xi^b \right),$ in Eq.\eqref{effective_action} is derived in Appendix \ref{Ap:Action_for_geodesic_devi}.
From this equation, the energy–momentum tensor $T^{\mu\nu}(x)=-2\frac{\delta S_\text{dev}}{\delta h_{\mu\nu}}$becomes
\begin{align}
    T^{\mu\nu}(x)&=m\int dt'  \xi^a \xi^b E_{0 a 0 b}^{\mu\nu} \delta^4(x-X(t')),
    \label{energy_mom_tensor1}\\
    E_{\rho\alpha\sigma\beta}^{\mu\nu}&=\frac{1}{2}[\partial_{\alpha}\partial_{\sigma}\delta_{\rho}^{(\mu}\delta_{\beta}^{\nu)} - \partial_{\alpha}\partial_{\beta}\delta_{\rho}^{(\mu}\delta_{\sigma}^{\nu)} - \partial_{\rho}\partial_{\sigma}\delta_{\alpha}^{(\mu}\delta_{\beta}^{\nu)} + \partial_{\beta}\partial_{\rho}\delta_{\alpha}^{(\mu}\delta_{\sigma}^{\nu)}],
    \label{E_differential}
\end{align}
where $\delta_{\alpha}^{(\mu}\delta_{\sigma}^{\nu)}=(\delta_{\alpha}^{\mu}\delta_{\sigma}^{\nu}+\delta_{\alpha}^{\nu}\delta_{\sigma}^{\mu})/2$.

The time evolution of the quantum geodesic deviation and the classical gravitational field is given as
\begin{align}
\rho[\xi^a_f,\underline{\xi}^a_f,h_{\mu\nu,f},t_f]=\frac{1}{N}\int^f \mathcal{D}\xi^a \mathcal{D}\underline{\xi}^a \mathcal{D}h_{\mu\nu} \delta[\partial^{\mu}(h_{\mu\nu}-\frac{1}{2}\eta_{\mu\nu}h)] e^{I_{\mathrm{CQ}}[\xi,\underline{\xi},h_{\mu\nu}]} \rho[\xi^a_i,\underline{\xi}^a_i,h_{\mu\nu,i},t_i],
\label{CQ_path_integral}
\end{align}
where the label $f$ at the upper limit of the path integral denotes the boundary condition  $\xi^a_f,\underline{\xi}^a_f,h_{\mu\nu,f}$, and the classical-quantum action, $I_{\mathrm{CQ}}$, is given as
\begin{align}
    I_{\mathrm{CQ}}
    &\notag=
    i(S_\text{tot}[\xi^a,h_{\mu\nu}]-S_\text{tot}[\underline{\xi}^a,h_{\mu\nu}]) \\
    &\notag \quad -\frac{1}{2}\int^{t_f}_{t_i} d^4x d^4y D_{\mu\nu\rho\sigma}(x,y)[T^{\mu\nu}(x)-\underline{T}^{\mu\nu}(x)] [T^{\rho\sigma}(y)-\underline{T}^{\rho\sigma}(y)] \\
    & \quad -\frac{1}{2}\int^{t_f}_{t_i} d^4x d^4y N^{-1}_{\mu\nu\rho\sigma}(x,y)[G^{\mu\nu(1)}(x)-4\pi G_N (T^{\mu\nu}(x)+\underline{T}^{\mu\nu}(x))] [G^{\rho\sigma(1)}(y)-4\pi G_N (T^{\rho\sigma}(y)+\underline{T}^{\rho\sigma}(y))].
    \label{CQ_action}
\end{align}
Here, the notation $\underline{T}^{\mu\nu}$ indicates that $\xi$ contained in $T^{\mu\nu}$ is replaced by $\underline{\xi}$.
The first line is a purely Schwinger--Keldysh path-integral term, while the second line represents a decoherence process for the quantum deviation, which is characterized by the decoherence kernel $D_{\mu\nu\rho\sigma}(x,y)$.
A larger value of $D_{\mu\nu\rho\sigma}(x,y)$ corresponds to a stronger suppression of the quantum spread 
$T^{\mu\nu}(x) - \underline{T}^{\mu\nu}(x)$ in the path integral. 
As a result, quantum coherence is more easily lost, i.e., stronger decoherence occurs. 
Therefore, $D_{\mu\nu\rho\sigma}(x,y)$ characterizes the strength of decoherence.
On the other hand, the third line represents a noise term of the gravitational field characterized by the noise kernel 
$N_{\mu\nu\rho\sigma}(x,y)$. 
Here, $G^{\mu\nu(1)}$ denotes the Einstein tensor expanded to first order in the metric perturbations $h_{\mu\nu}$. 
In the regime where $N_{\mu\nu\rho\sigma}(x,y)$ becomes small, 
the quantity $G^{\mu\nu(1)}(x) - 4\pi G_N \bigl(T^{\mu\nu}(x) + \underline{T}^{\mu\nu}(x)\bigr)$ approaches zero in the path integral. 
In this limit, when the quantum deviation becomes classical so that 
$T^{\mu\nu}(x) = \underline{T}^{\mu\nu}(x)$, 
the deterministic Einstein equations is recovered. 
Therefore, $N_{\mu\nu\rho\sigma}(x,y)$ characterizes the magnitude of the noise inherent in the gravitational field.

For the action given in Eq.~\eqref{CQ_action}, the decoherence-diffusion trade-off relation in Eq.~\eqref{trade_off} takes the form
\begin{equation}
    DN \ge (4\pi G_N)^2.
    \label{our_tradeoff}
\end{equation}
This is a matrix inequality about the matrices $D_{\mu\nu\rho\sigma}(x,y)$ and $N_{\mu\nu\rho\sigma}(x,y)$. 

Now we replace $\xi^a$ with $L^a+\xi^a$ and assume that  $\xi^a$ is a small displacement from the mean separation $L^a$. 
From Eq.~\eqref{CQ_path_integral}, the Langevin equation for the geodesic deviation,
\begin{align}
    m\frac{d^2 \xi^a}{dt^2} - \zeta^a(t)=0,
    \label{Langevin_equation}
\end{align}
is obtained by assuming that the initial conditions $h_{\mu \nu} (t_i,\bm{x})=0=\dot{h}_{\mu\nu} (t_i,\bm{x})$ and neglecting gravitational waves radiated from the motion of deviation.
The stochastic force $\zeta_a (t)=\delta_{ab} \zeta^b(t)$ satisfies
\begin{align}
\braket{\zeta_a(t)}&=0, \quad \braket{\zeta_a(t) \zeta_b(t')}= \left[ \Delta^D_{cabd}(t,t') + \Delta^N_{cabd}(t,t') \right] L^cL^d,\label{two_point_correlation} 
\end{align}
with 
\begin{align}
\Delta^D_{abcd}(t,t')&= 4m^2 E_{0a0b}^{x,\mu\nu} E_{0c0d}^{y,\rho\sigma} D_{\mu\nu\rho\sigma}(x,y)|_{x^\mu=X^\mu(t), y^\mu=X^\mu(t')}, 
\label{DeltaD}
\\
\Delta^N_{abcd}(t,t')&= 16m^2 \int^t_{t_i} d^4 z  \int^{t'}_{t_i} d^4 w  E_{0a0b}^{x,\mu\nu} G_R(x-z) E_{0c0d}^{y,\rho\sigma} G_R(y-w)|_{x^\mu=X^\mu(t), y^\mu=X^\mu(t')} \notag\\
&\quad \times (\delta^{\alpha}_{\mu} \delta^{\beta}_{\nu} - \frac{1}{2}\eta_{\mu\nu}\eta^{\alpha\beta}) (\delta^{\lambda}_{\rho} \delta^{\kappa}_{\sigma} - \frac{1}{2}\eta_{\rho\sigma}\eta^{\lambda\kappa}) N_{\alpha \beta \lambda \kappa}(z,w)
\label{DeltaN},
\end{align}
where $t_i$ is an initial time, and the differential operators $E_{0a0b}^{x,\mu\nu}$ and $E_{0c0d}^{y,\rho\sigma}$ defined in Eq.~\eqref{E_differential} act on $x$ and $y$. The $G_R$ is the retarded Green's function. 
See Appendix~\ref{Derivation of Langevan equation} for the detail calculation.
Eq.~\eqref{two_point_correlation} implies that the decoherence effect of the deviation acts directly on itself as a term $\Delta^{D}_{abcd}$, 
while the noise inherent in the gravitational field also exerts a term $\Delta^{N}_{abcd}$ on the deviation. 
These terms $\Delta^{D}_{abcd}$ and $\Delta^{N}_{abcd}$ originated from the decoherence kernel $D_{\mu\nu\rho\sigma}(x,y)$ and the noise kernel $N_{\mu\nu\rho\sigma}(x,y)$, respectively, are now in a trade-off relation.
In the next section, we focus on the two-point correlation of the stochastic force $\zeta_a$, Eq.~\eqref{two_point_correlation}, and analyze the characteristics exhibited by its spectra.

\section{Strain Spectra for CQ models}
\label{sec:Noise_spectra}

In this section, we compute the power spectral density expected from the fluctuations of the geodesic deviation using the expressions derived so far. 
In gravitational-wave experiments, the dimensionless quantity known as the strain is commonly used as an observable, 
and accordingly we also present our final results in terms of the strain.
If $L$ denotes the mean separation between two test masses and $\Delta L(t)$ the relative change in their separation, 
the strain $h(t)$ is defined as
\begin{equation}
    h(t) \equiv \frac{\Delta L(t)}{L}.
\end{equation}
That is, the strain measures the magnitude of the relative stretching and squeezing of spacetime caused by gravitational fields, 
and it is the fundamental observable measured in gravitational-wave experiments.

With these definitions in mind, starting from the correlation function of the geodesic deviation $\xi^a(t)$ obtained by Eq.~\eqref{two_point_correlation}, 
we define the power spectral density $\left(S^{h}_{x}\right)^2$ as
\begin{equation}
    \left(S^{h}_{x}\right)^2
    =
    \frac{1}{m^2 L^2 \omega^4}
    \int dt \, e^{i \omega t}
    \braket{\zeta_x(t)\zeta_x(0)}.
    \label{root_PSD}
\end{equation}
Here, $m$ is the mass in Eq.\eqref{Langevin_equation}, $L$ is the mean separation length between $M$ and $m$, and $\omega$ is the angular frequency. 
In Eq.~\eqref{two_point_correlation}, we take the indices $a,b$ to be along the $x$ direction $(a=b=x)$ and set $L^{c,d} = [L,0,0]^\text{T}$. 
This choice corresponds to evaluating how the fluctuations of the geodesic deviation in the direction of $[L,0,0]^\text{T}$
are correlated between different times. To evaluate the power spectral density, the initial time $t_i$ is taken to the limit $t_i \rightarrow -\infty$. 
Substituting the correlation function of $\zeta_b$, Eq.~\eqref{two_point_correlation}, into the power spectral density, we get 
\begin{equation}
(S^h_x)^2=(S^D_x)^2+(S^N_x)^2,
\label{SD+SN}
\end{equation}
where
\begin{align}
    \left(S^{D}_{x}\right)^2
    =
    \frac{1}{m^2 \omega^4}
    \int dt e^{i\omega t}\Delta^D_{xxxx}(t,0), 
    \quad  
    \left(S^{N}_{x}\right)^2
    =
    \frac{1}{m^2 \omega^4}
    \int dt e^{i\omega t}\Delta^N_{xxxx}(t,0). 
    \label{SDN}
\end{align}
Each contribution comes from the decoherence of the quantum deviation and the noise of the classical gravitational field, respectively.
In the plots presented in the following sections, we use the square root of the power spectral density defined above, $S^{h}_{x}$, which is called the strain spectrum in this paper.

\subsection{Classical-quantum model proposed by Oppenheim \emph{et al}.}
\label{section:Oppenheim original model}
In Oppenheim \emph{et al}. original model proposed in \cite{grudka2024Renormalisation}, the simple example of $D_{\mu\nu\rho\sigma}(x,y)$ and $N_{\mu\nu\rho\sigma}(x,y)$ that determine the correlation $\Delta^D_{abcd}(t,t')$ and $\Delta^N_{abcd}(t,t')$ are given by
\begin{align}
    D_{\mu\nu\rho\sigma}(x,y)
    &=\frac{D_0^{\mathrm{\,org}}}{8}(\eta_{\mu\rho}\eta_{\nu\sigma} + \eta_{\mu\sigma}\eta_{\nu\rho} - 2\beta \eta_{\mu\nu}\eta_{\rho\sigma}) \delta^4(x-y),
    \label{Op_Deco} \\
    N_{\mu\nu\rho\sigma}(x,y)
    &=2N^{\mathrm{\,ori}}_0  (\eta_{\mu\rho}\eta_{\nu\sigma} + \eta_{\mu\sigma}\eta_{\nu\rho} +\frac{2\beta}{1-4\beta} \eta_{\mu\nu}\eta_{\rho\sigma}) \delta^4(x-y),
    \label{Op_Noise}
\end{align}
where $\beta$ is a parameter that takes values between $0$ and $1$, $D_0^{\mathrm{\,org}}$ and $N_0^{\mathrm{\,org}}$ are non-negative constants.
These two constants satisfy the tradeoff relation $D_0^{\mathrm{\,org}} N_0^{\mathrm{\,org}} \ge (4\pi G_N)^2$, which follows from Eq.~\eqref{our_tradeoff} for $D_{\mu\nu\rho\sigma}(x,y)$ and $N_{\mu\nu\rho\sigma}(x,y)$.
Here, one can choose  $N_0^{\mathrm{\,org}} = (4\pi G_N)^2 / D_0^{\mathrm{\,org}}$.
This saturates the trade-off relation, and the other choices of $N_0^{\mathrm{\,org}}$ would lead to larger fluctuations of deviation.

Using Eqs.~\eqref{Op_Deco} and~\eqref{Op_Noise}, and taking the limit $t_i \rightarrow -\infty$, the power spectral density defined in Eq.~\eqref{root_PSD} can be calculated as
\begin{equation}
    (S^h_{x,\text{org}})^2 = (S^D_{x,\text{org}})^2 + (S^N_{x,\text{org}})^2,
    \label{Spectrum_Oppenheim_original}
\end{equation}
where $S^D_{x,\text{org}}$ and $S^N_{x,\text{org}}$ are 
\begin{align}
    (S^D_{x,\text{org}})^2&=
    \frac{D_0^{\mathrm{\,org}}}{\pi^2}(1-\beta) \left( \frac{1}{3L^3} - \frac{2}{15L^5\omega^2} + \frac{1}{35L^7\omega^4} \right),
    \label{D_Spectrum_Oppenheim_original}\\ 
    (S^N_{x,\text{org}})^2&\notag = \frac{512}{ D_0^{\mathrm{\,org}} m_p^4} \frac{1-3\beta}{1-4\beta} \left[ \frac{(\omega+i\epsilon)\mathrm{Arccot}[L(\epsilon-i\omega)] + (\omega-i\epsilon)\mathrm{Arccot}[L(\epsilon+i\omega)]}{4\epsilon\omega}\right. \\\notag
    &\qquad -   \frac{4\omega\epsilon-iL(\epsilon-i\omega)^3\mathrm{Arccot}[L(\epsilon-i\omega)] + iL(\epsilon+i\omega)^3\mathrm{Arccot}[L(\epsilon+i\omega)]}{6 L\epsilon\omega^3} \\
    &\qquad \left. +  \left( \frac{1}{15L^3 \omega^4} + \frac{2(\omega^2 - \epsilon^2)}{5L\omega^4} + \frac{(\omega+i\epsilon)^5\mathrm{Arccot}[L(\epsilon-i\omega)] + (\omega-i\epsilon)^5\mathrm{Arccot}[L(\epsilon+i\omega)]}{20\epsilon\omega^5} \right) \right],
    \label{N_Spectrum_Oppenheim_original}
\end{align}
where $G_N=1/m^2_p$.
We also introduced the UV/IR cutoffs so that incoming gravitational fields whose wavelengths are smaller than the separation size $L$, and the time over which the noise accumulates is limited to the current age of the universe, $1/\epsilon$, respectively\footnote{Here we assume that the noise in the gravitational field has been present throughout the evolution of the universe; however, for simplicity, the contribution from cosmic expansion is neglected.
In order to clarify the contribution, it would be necessary to construct a CQ model on an expanding spacetime. This lies beyond the scope of the present paper and will not be addressed here.}.
This parameter $\epsilon$ comes from the fact that, in the present calculation, the retarded Green's function
\begin{equation}
  G_R(x-y)=\int_{\mathbb{R}^4}\frac{d^4k}{(2\pi)^4} \frac{e^{ik_{\mu}(x^{\mu}-y^{\mu})}}{-(k^0+i\epsilon)^2+\boldsymbol{k}^2},
  \label{R_Green_function}
\end{equation}
was introduced in order to make the integrals convergent
\footnote{In Ref.~\cite{oppenheim2025diffusionstochastickleingordonequation}, the two-point correlation function of a field with a stochastic source is evaluated.
Two approaches are compared: one based on the retarded Green's function with an $\epsilon$ prescription,
and another assuming a finite time interval.
Although the results obtained from these analyses do not completely coincide,
they are expected to yield the same two-point correlation function in the long-time limit.
For this reason, we adopt the Green's function method in the following.}.
As will be discussed in the next section, if the gravitational fields possess scale-free noise that does not depend on environmental degrees of freedom, the theory without these two types of cutoffs is divergent, which is a distinctive feature of this model. 
However, if one takes into account the dissipative term in the Langevin equation~\eqref{Langevin_equation}, 
which is neglected in the present analysis, such divergences may be avoided.

\begin{figure}[htbp]
  \centering  \includegraphics[width=1.0\linewidth]{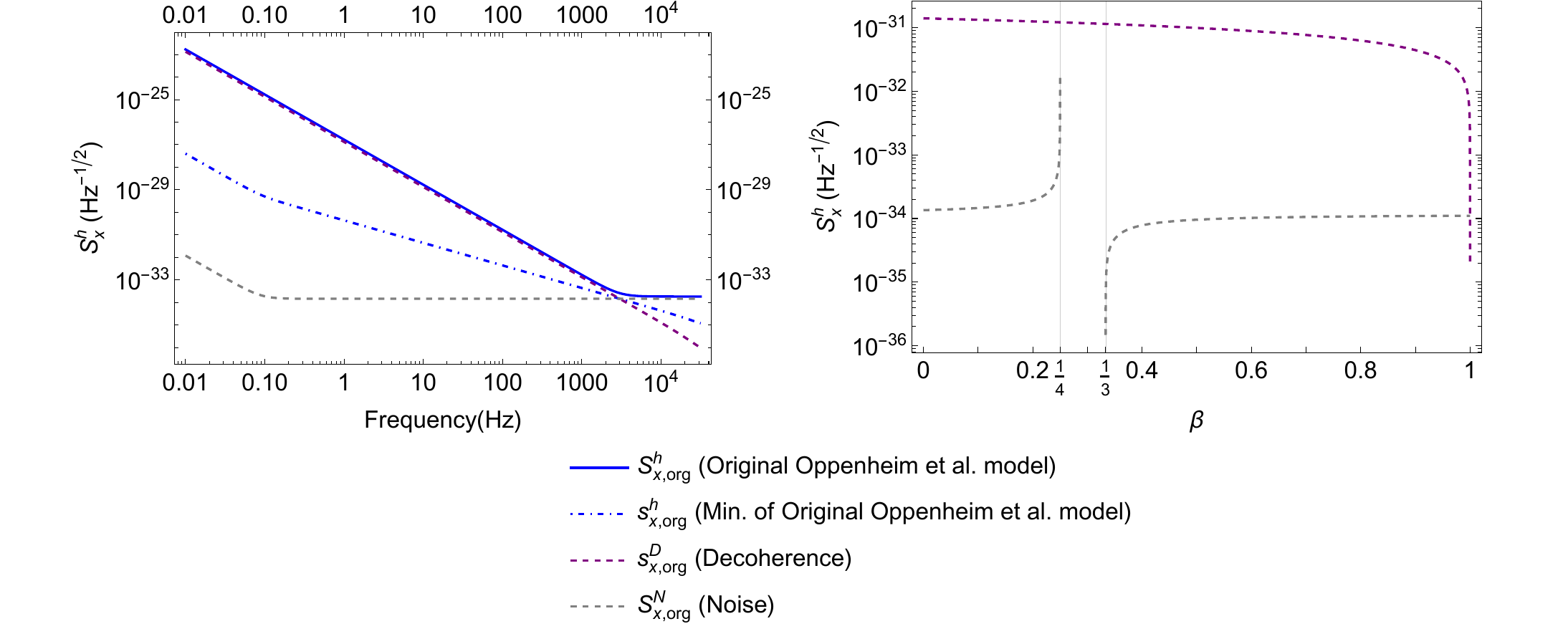}
  \caption{Upper panel: the strain spectra of the original Oppenheim \emph{et al}. model $S^h_{x,\text{org}}$ and its minimum $s^h_{x,\text{org}}$. The blue solid line shows $S^h_{x,\text{org}}$ for $\beta = 0.1$, $D_0^{\mathrm{\,org}} = 10^{-85}\,\mathrm{Hz}^{-4}$, 
$L = 4\,\mathrm{km}$, and $\epsilon = 10^{-18}\,\mathrm{Hz}$. 
The parameter $1/\epsilon$ corresponds to the age of the universe. 
The blue dot-dashed line shows the minimum of the spectrum, which is independent of $D_0^{\mathrm{\,org}}$; the remaining parameters are fixed to the same values as in $S^h_{x,\text{org}}$.
Further, $S^D_{x,\text{org}}$ contributed from $\Delta^D_{abcd}$ (shown in dashed purple) and $S^N_{x,\text{org}}$ derived from $\Delta^N_{abcd}$ (shown in dashed gray) are plotted separately. The parameters used in the plot are the same as those used for the blue solid line.
Lower panel: the $\beta$ dependence of $S^h_{x,\text{org}}$ is shown for $D_0^{\mathrm{\,org}} = 10^{-85}\,\mathrm{Hz}^{-4}$, 
$L = 4\,\mathrm{km}$, $\epsilon = 10^{-18}\,\mathrm{Hz}$ and $\omega = 100\,\mathrm{Hz}$. As observed in Eq.~\eqref{N_Spectrum_Oppenheim_original}, $S^N_{x,\text{org}}$ diverges at $\beta = 1/4$ and vanishes at $\beta = 1/3$. In the range $1/4 < \beta < 1/3$, it becomes complex.
Except for a region of $\beta$ that yields such pathological values, the strain spectrum remains nearly unchanged. Therefore, we will continue to adopt $\beta = 0.1$ in the following analysis.}
  \label{fig:plot1_op}
\end{figure}

The left panel of Fig.~\ref{fig:plot1_op} shows the plot of the spectrum of the strain $S^h_{x,\text{org}}$ with blue solid line.
The strain spectrum $S^h_{x,\text{org}}$ in \eqref{Spectrum_Oppenheim_original} depends on the parameter $D_0^{\mathrm{\,org}}$ characterizing the Oppenheim \emph{et al}. model. 
However, by taking the arithmetic and geometric means of $(S^D_{x,\text{org}})^2$ and $(S^N_{x,\text{org}})^2$, one obtains the minimum strain spectrum
\begin{align}
    s^h_{x,\text{org}} = 2 \sqrt{(S^D_{x,\text{org}})^2 \times (S^N_{x,\text{org}})^2},
    \label{mini_Spectrum_Oppenheim_original}
\end{align}
which is independent of $D_0^{\mathrm{\,org}}$ since it cancels out.
This minimal strain is plotted as blue dot-dashed line in the left panel of Fig.~\ref{fig:plot1_op}.

In Fig.~\ref{fig:plot1_op}, for any choice of $D_0^{\mathrm{\,org}}$, the spectrum never falls below the minimum line. This indicates that, regardless of the parameter values, a minimum fluctuation always exists. Although the strain itself is small, its magnitude increases with $D_0^{\mathrm{\,org}}$. Comparing the magnitude of strain spectrum with the sensitivity obtained from experiments, we can constraint the parameter region of $D_0^{\mathrm{\,org}}$. This point will be examined in later sections.

The bending of the solid blue line in the left panel is interpreted as a manifestation of the decoherence--diffusion trade-off relation. 
In the left panel, the contribution of $S^D_{x,\text{org}}$ from the decoherence kernel $D_{\mu\nu\rho\sigma}$ is shown in dashed purple, while the contribution of $S^N_{x,\text{org}}$ from the noise kernel $N_{\mu\nu\rho\sigma}$ is shown in dashed gray. As can be seen from Eq.~\eqref{Spectrum_Oppenheim_original} and the left panel of Fig.~\ref{fig:plot1_op}, the contribution from the decoherence kernel dominates in the low-frequency region, whereas the contribution from the noise becomes dominant in the high-frequency region.
In the present analysis, the parameter $1/\epsilon$ is taken to be of the order of the age of the universe; however, if one considers a sufficiently far future, namely if $\epsilon$ becomes sufficiently small, the spectra diverges. This implies that the original model suffers from the problem that the observable spectra diverges in the far-future limit.

The right panel of Fig.~\ref{fig:plot1_op} shows the $\beta$ dependence of the contributions of $S^D_{x,\text{org}}$ and $S^N_{x,\text{org}}$ to $S^h_{x,\text{org}}$. As can also be inferred from Eq.~\eqref{N_Spectrum_Oppenheim_original}, when $\beta \in \left[\frac{1}{4},\frac{1}{3}\right]$, $S^N_{x,\text{org}}$ exhibits a singular feature. 
When $\beta = \tfrac{1}{4}$, the noise contribution $S^N_{x,\text{org}}$ diverges. For $\beta = \tfrac{1}{3}$, it vanishes, and the spectrum is determined solely by the decoherence contribution $S^D_{x,\text{org}}$. 
For $\tfrac{1}{4} < \beta < \tfrac{1}{3}$, $(S^N_{x,\text{org}})^2$ is negative and the strain spectrum $S^h_{x,\text{org}} $becomes complex. This feature reflects the fact that the noise kernel is not positive semidefinite. To keep the positive semidefiniteness and get a finite result in our analysis, we should consider the region $\beta <\tfrac{1}{4}$ or $ \beta \geq \tfrac{1}{3}$.

\subsection{Modified Oppenheim \emph{et al}. model consistent with Bianchi identities}
\label{sec:CQwithEin}

In the CQ path integral \eqref{CQ_path_integral} with the CQ action \eqref{CQ_action}, the gravitational field stochastically behaves due to the presence of the noise kernel $N_{\mu\nu\rho\sigma}$. 
We consider that the noise kernel $N_{\mu\nu\rho\sigma}(x,y)$ given in Eq.~\eqref{Op_Noise} should be chosen to be consistent with the Bianchi identities. 
The Einstein equations of the gravitational field takes the form \begin{align}
    G^{(1)}_{\mu\nu}=\chi_{\mu\nu},
    \label{noise_Einstein_Langevin}
\end{align}
where $\chi_{\mu \nu}$ is the stochastic source satisfying 
\begin{align}
    \braket{\chi_{\mu\nu}(x)}&=0,\quad \braket{\chi_{\mu\nu} (x) \chi_{\rho\sigma}(y)}=N_{\mu\nu\rho\sigma} (x,y).
    \label{eq:chi}
\end{align}
In the above Einstein equations, for simplicity, the energy-momentum tensor \eqref{energy_mom_tensor1} is neglected by assuming that the mass $m$ in \eqref{energy_mom_tensor1} is sufficiently small. 
Eq.\eqref{noise_Einstein_Langevin} is called the \textit{(linearized) Einstein–Langevin equation}.
We can compute the spacetime divergence of the Einstein-Langevin equation as \footnote{Now, since we consider the linear perturbation in $h_{\mu\nu}$ around the flat spacetime, the covariant derivative coincides with the partial derivative ($\nabla_{\mu}=\partial_{\mu}$).}:
\begin{align}
    \partial^{\mu} G^{(1)}_{\mu\nu}=\partial^{\mu}\chi_{\mu\nu}.
\end{align}
The left-hand side should be zero because of the Bianchi identity. 
If we only observe the statistical behavior of the gravitational field, the Bianchi identity suggests that the stochastic source $\chi_{\mu \nu}$ follows
\begin{align}
    \braket{\partial^{\mu} \chi_{\mu\nu}(x)}=0, \quad \braket{ \partial^{\mu} \chi_{\mu\nu} (x) \chi_{\rho\sigma}(y) }= \partial^{\mu}N_{\mu\nu\rho\sigma}(x,y)=0.
    \label{Noise_Bianchi}
\end{align}
The first equation of Eqs.~\eqref{Noise_Bianchi}, $\braket{\partial^{\mu} \chi_{\mu\nu}(x)}=0$, is automatically satisfied.
On the other hand, the noise kernel $N_{\mu\nu\rho\sigma} (x,y)$ does not satisfy the second equation of Eqs.~\eqref{Noise_Bianchi}, $\partial^{\mu}N_{\mu\nu\rho\sigma}(x,y)=0$, since the noise kernel is proportional to $\delta^4 (x-y)$ as given in Eq.~\eqref{Op_Noise}. \footnote{Exactly speaking, the stochastic source $\chi_{\mu\nu}$ with the noise kernel $N_{\mu\nu\rho\sigma}(x,y) \propto \delta^4(x-y)$ is a white noise, and the temporal and spatial derivatives of $\chi_{\mu\nu}$ diverge and are ill-defined. We need to control such a divergence to keep the consistency with the Bianchi identities. In this paper, we do not address this problem.}. 
Even if we have the energy-momentum tensor $T_{\mu\nu}$ of Eq.~\eqref{Op_Noise}, it does not affect the above discussion because we can check the conservation law $\partial^\mu T_{\mu\nu}(x)=0$. 
Regarding the consistency with the Bianchi identity, in Ref. \cite{Freidel2026geometric}, the general form of the Lorentz-invariant noise kernel $N_{\mu\nu\rho \sigma} (x,y)$ following $\partial^{\mu}N_{\mu\nu\rho\sigma}(x,y)=0$ was discussed.

\begin{figure}[H]
  \centering  \includegraphics[width=0.9\linewidth]{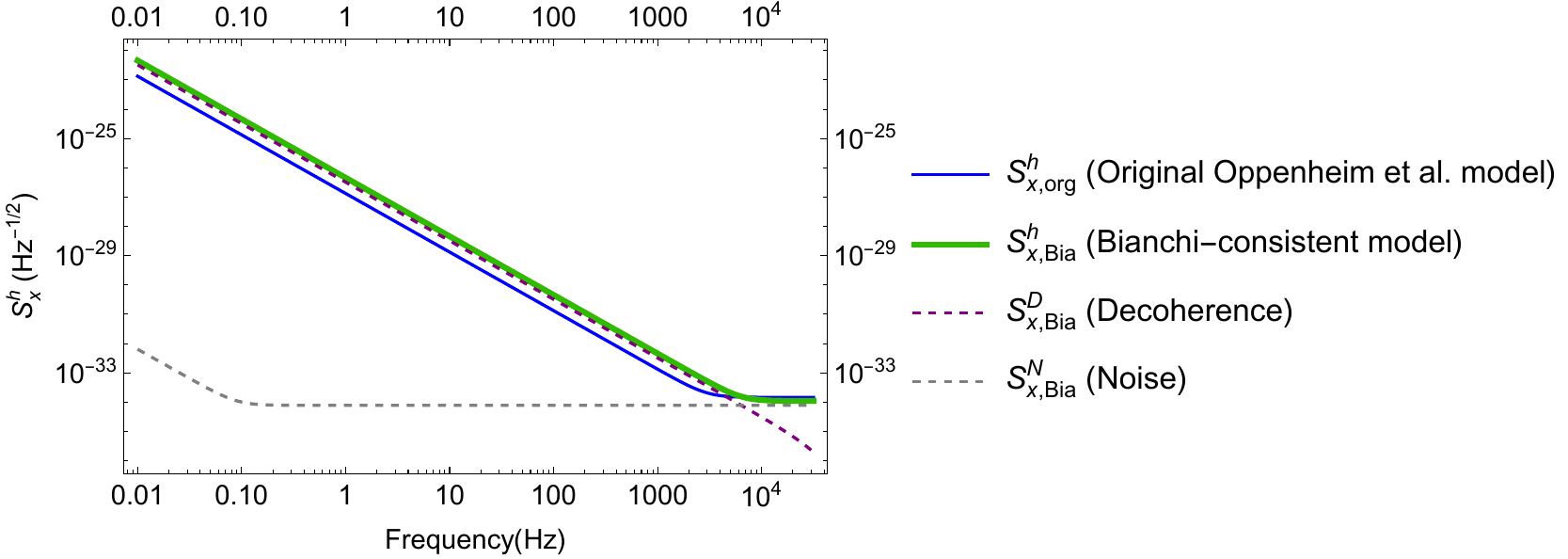}
  \caption{Strain spectrum $S^h_{x,\text{Bia}}$ given in Eq.~\eqref{Spectrum_white_Ein} for the Bianchi-consistent model. 
The green solid line shows $S^h_{x,\text{Bia}}$ plotted with $D_0^{\mathrm{\,Bia}} = 10^{-85}\,\mathrm{Hz}^{-4}$, 
$L = 4\,\mathrm{km}$, and $\epsilon = 10^{-18}\,\mathrm{Hz}$. 
The blue solid line corresponding to the strain $S^h_{x,\text{org}}$ obtained in the original Oppenheim \emph{et al}. model is plotted with $\beta = 0.1$, while all other parameters are set to be the same as those used for the green line.
In Eq.~\eqref{Spectrum_white_Ein}, the contributions from decoherence $S^D_{x,\text{Bia}}$ (shown in dashed purple) and from noise $S^N_{x,\text{Bia}}$ (shown in dashed gray) are presented separately. The parameters used in the plot are the same as green solid line.
}
  \label{fig:plot3_OP}
\end{figure}

Here, we propose a noise kernel consistent with the Bianchi identities, that is, satisfying $\partial^{\mu}N_{\mu\nu\rho\sigma}(x,y)=0$. 
As a simple modified version of Oppenheim \emph{et al}. model, we consider the following scale-free noise kernel,
\begin{align}
    N_{\mu\nu\rho\sigma}(x,y)
    &=\frac{(4\pi G_N)^2}{D_0^{\mathrm{\,Bia}}}\int \frac{d^4p}{(2\pi)^4} e^{ip^\mu (x_\mu -y_\mu)}  \mathcal{P}_{\mu\nu\rho\sigma},
    \label{Ein_Noise}
\end{align}
where 
\begin{equation}
    \mathcal{P}_{\mu\nu\rho\sigma} = \mathcal{P}_{\mu\rho}\mathcal{P}_{\sigma\nu} + \mathcal{P}_{\mu\sigma}\mathcal{P}_{\rho\nu} -\frac{2}{3}\mathcal{P}_{\mu\nu}\mathcal{P}_{\rho\sigma}, \quad \mathcal{P}_{\mu\nu} = \eta_{\mu\nu} - \frac{p_{\mu} p_{\nu}}{p^2}.
    \label{def_P}
\end{equation}
The projection tensor $\mathcal{P}_{\mu\nu}$ satisfies $p^\mu \mathcal{P}_{\mu\nu}=0$, hence we have $\partial^\mu N_{\mu\nu\rho \sigma}(x,y)=0$, and the noise kernel is consistent with the Bianchi identities.
We also adopt the decoherence kernel,
\begin{align}
    D_{\mu\nu\rho\sigma}(x,y)
    &= D_0^{\mathrm{\,Bia}} \int \frac{d^4p}{(2\pi)^4} e^{ip^\mu (x_\mu -y_\mu)}  \mathcal{P}_{\mu\nu\rho\sigma}. 
    \label{Ein_Deco}
\end{align}
These kernels are constructed to satisfy the trade-off relation $D_0^{\mathrm{\,Bia}} N_0^{\mathrm{\,Bia}} \ge (4\pi G_N)^2$.
However, we do not verify in this paper whether the
trade-off relation \eqref{our_tradeoff} also holds for non-local kernels that are not proportional
to $\delta^4(x-y)$. Since our
primary interest lies in exploring the consequences of imposing
trade-off relation \eqref{our_tradeoff}, we do not pursue this issue further here.
The noise and decoherence kernels give the power spectral density,
\begin{equation}
    (S^h_{x,\text{Bia}})^2 = (S^D_{x,\text{Bia}})^2 + (S^N_{x,\text{Bia}})^2,
    \label{Spectrum_white_Ein}
\end{equation}
where $S^D_{x,\text{Bia}}$ and $S^N_{x,\text{Bia}}$ are given by
\begin{align}
    (S^D_{x,\text{Bia}})^2
    &
    = \frac{8D_0^{\mathrm{\,Bia}}}{\pi^2}\left( \frac{1}{9L^3} -\frac{2}{45L^5 \omega^2} +\frac{1}{105L^7\omega^4} \right),
    \label{D_Spectrum_white_Ein}\\ \notag
    (S^N_{x,\text{Bia}})^2&= \frac{128}{D_0^{\mathrm{\,Bia}} m_p^4} \left[  \frac{(\omega +i\epsilon)\mathrm{Arccot}[L(\epsilon-i\omega)] + (\omega -i\epsilon)\mathrm{Arccot}[L(\epsilon+i\omega)]}{3\epsilon \omega}\right. \\\notag
    &\qquad - 2 \frac{4\epsilon \omega - iL (\epsilon - i\omega)^3 \mathrm{Arccot}[L(\epsilon-i\omega)] + iL (\epsilon + i\omega)^3 \mathrm{Arccot}[L(\epsilon+i\omega)]}{9L\epsilon \omega^3} \\
    &\qquad \left. + \frac{4}{15}  \left( \frac{1}{3L^3 \omega^4} + \frac{2(\omega^2 - \epsilon^2)}{L\omega^4} + \frac{(\omega + i\epsilon)^5 \mathrm{Arccot}[L(\epsilon-i\omega)] +  (\omega - i\epsilon)^5 \mathrm{Arccot}[L(\epsilon+i\omega)]}{4\epsilon \omega^5} \right) \right].
    \label{N_Spectrum_white_Ein}
\end{align}
As in the original model, this model requires two UV/IR cutoff parameters, $L$ and $\epsilon$, respectively. 
For comparison, the resulting strain spectrum $S^h_{x,\text{Bia}}$ is plotted together with that of the original model as a green solid line, as shown in Fig.~\ref{fig:plot3_OP}.
The mean separation length $L$ and the age of the universe $1/\epsilon$ used in the plot are chosen to be identical to Fig.\ref{fig:plot1_op}.
We introduced a scale-free noise kernel, as in the original model, and is consistent with the Bianchi identities. However, in practice it appears that this modification makes little difference when evaluating the fluctuations of the geodesic deviation.

\subsection{Classical-quantum model with environment-induced noise}
\label{sec:CQwithEnv}

An unclear point in the classical-quantum models discussed so far is the origin of noise attributed to the classical gravitational field. In the original formulation by Oppenheim \emph{et al.}, such a noise is introduced axiomatically as an intrinsic property of the gravitational field, without a detailed discussion of their physical origin. 
While this assumption is operationally effective at the level of a phenomenological model, it appears somewhat unnatural from the standpoint of physical intuition, as the origin of noise remains obscure.

Motivated by this concern, we propose that the gravitational field is regarded as fundamentally quantum mechanical, interacting with additional quantum degrees of freedom that play a role of an environment. Upon coarse-graining these environmental degrees of freedom, i.e.\ tracing them out, the gravitational field exhibits an effectively classical and stochastic behavior. In this picture, the noise appearing in the classical gravitational field is induced by the environment.

\begin{figure}[H]
  \centering
  \includegraphics[width=0.9\linewidth]{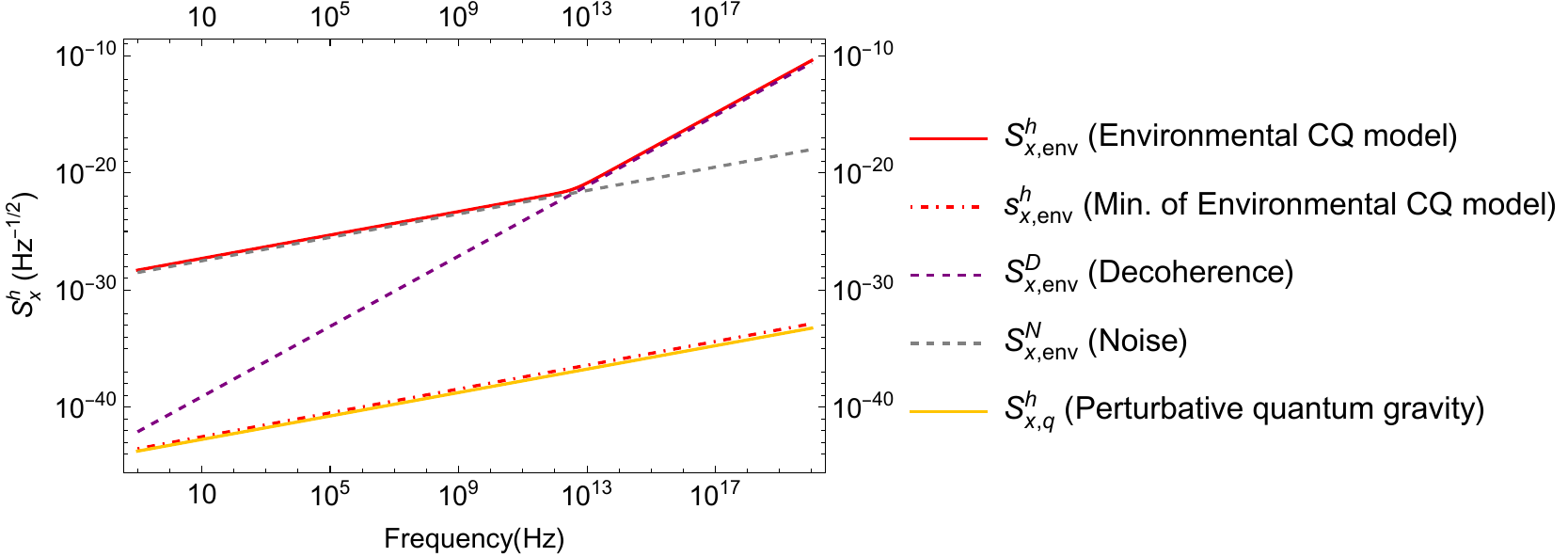}
  \caption{Spectra predicted from the environmental CQ model. 
The red solid line shows $S^h_{x,\text{env}}$ given by Eq.~\eqref{Spectrum_Environmental_Oppenheim} plotted with $D_0^{\mathrm{\,env}} = 10^{-80}\,\mathrm{Hz}^{-4}$, and 
$\mu = 10^{-18}\,\mathrm{Hz}$.
This plot does not depend on $\beta, L, \epsilon$.
The $\mu$ corresponds to a scale given by the inverse of the age of the universe.
The red dot-dashed line represents the minimum spectrum, $s^h_{x,\text{env}}$, which is independent of $D_0^{\mathrm{\,env}}$.  
The yellow line corresponds to the strain predicted from perturbative quantum gravity \eqref{PSD_QG}.
In Eq.~\eqref{Spectrum_Environmental_Oppenheim}, the contributions $S^D_{x,\text{env}}$ from the decoherence kernel (shown in dashed purple) and $S^N_{x,\text{env}}$ from the noise kernel (shown in dashed gray) are plotted separately. The parameters used in the plot are the same as the red solid line.
}
  \label{fig:plot2_OP}
\end{figure}

A characteristic feature of the environment-induced noise is that the noise spectrum has the energy scale of environment.
For simplicity, we phenomenologically model the noise kernel as \cite{PhysRevD.49.1861,PhysRevD.53.1927,MARTIN1999113,PhysRevD.61.124024} 
\begin{align}
    N_{\mu\nu\rho\sigma}(x,y)
    &=
    \int \frac{d^4p}{(2\pi)^4} N(p) e^{ip^\mu (x_\mu -y_\mu)} \theta(-p^2-4\mu^2) \mathcal{P}_{\mu\nu\rho\sigma},
    \label{EOp_Noise}
\end{align}
where $N(p)$ is a function of the four-momentum $p^\mu$, and $\mu$ represents the energy scale of environment. The step function $\theta(-p^2-4\mu^2)$ in the noise kernel means that the gravitational field with an energy larger than $2\mu$ is induced by the environmental noise. This kind of noise kernel was discussed in stochastic gravity approaches \cite{Hu2008_Stochastic_Gravity,PhysRevD.53.1927,PhysRevD.49.1861}. 
Specifically, the step function $\theta(-p^2-4\mu^2)$ and the tensor $\mathcal{P}_{\mu\nu\rho\sigma}$ were observed for the coupled model of a quantized gravitational field and a conformal scalar field with a small mass\,\cite{PhysRevD.61.124024}. Based on the above discussion in Sec.\ref{sec:CQwithEin}, the noise kernel is consistent with Bianchi identities in the sense that $\partial^\mu N_{\mu\nu\rho \sigma}(x,y)=0$ holds.
We also assume the following decoherence kernel, 
\begin{align}
    D_{\mu\nu\rho\sigma}(x,y)
    &= 
    \int \frac{d^4p}{(2\pi)^4} D(p) e^{ip^\mu (x_\mu -y_\mu)} \theta(-p^2-4\mu^2) \mathcal{P}_{\mu\nu\rho\sigma},
    \label{EOp_Deco} 
\end{align}
where 
$D(p)$ is a function of the four-momentum $p^\mu$. 
This decoherence kernel is simply chosen to have a form similar to the noise kernel.
From the trade-off relation \eqref{our_tradeoff}, the functions $N(p)$ and $D(p)$ should follow
\footnote{In stochastic gravity approaches \cite{Hu2008_Stochastic_Gravity,PhysRevD.53.1927,PhysRevD.49.1861}, the decoherence kernel \eqref{EOp_Deco} and the trade-off relation \eqref{our_tradeoff} are not generally derived. Here, we introduce them in order to embed the model into the CQ framework.}
\begin{align}
    D(p)N(p) \geq (4\pi G_N)^2.
    \label{NDp} 
\end{align}
A simple choice of them is
\begin{align}
    D(p)=D_0^{\mathrm{\,env}}, \quad N(p)=\frac{(4\pi G_N)^2}{D_0^{\mathrm{\,env}}},
    \label{eq:NDp} 
\end{align} 
and then the power spectral density is given by 
\begin{equation}
    (S^h_{x,\text{env}})^2 = (S^D_{x,\text{env}})^2 + (S^N_{x,\text{env}})^2,
    \label{Spectrum_Environmental_Oppenheim}
\end{equation}
where $S^D_{x,\text{env}}$ and $S^N_{x,\text{env}}$ are given by
\begin{align}
    (S^D_{x,\text{env}})^2
    &=\theta (\omega - 2\mu) \frac{64D_0^{\mathrm{\,env}}  (3\omega^4 +4\mu^2\omega^2 +6\mu^4) (\omega^2 - 4\mu^2)^{\frac{3}{2}} }{315 \pi^2\omega^4},
    \label{D_Spectrum_Environmental_Oppenheim} \\
    (S^N_{x,\text{env}})^2
    &=\theta (\omega - 2\mu) \frac{512(\omega^2+\mu^2) (\omega^2 - 4\mu^2)^{\frac{3}{2}} }{45D_0^{\mathrm{\,env}} m_p^4 \mu^2\omega^4}.
    \label{N_Spectrum_Environmental_Oppenheim}
\end{align}
As in the previous section, one can determine the minimal fluctuation independent of the parameter $D_0^{\mathrm{\,env}}$. In the present case, such a minimal fluctuation can be evaluated at the level of the two-point correlation function \eqref{two_point_correlation},
\begin{align}
\Delta^D_{abcd} + \Delta^N_{abcd}
&\ge \Delta^\text{min}_{abcd}=\frac{64 \pi m^2}{m_p^2}  E_{0a0b}^{x,\mu\nu} E_{0c0d}^{y,\rho\sigma} \int \frac{d^4p}{(2\pi)^4} e^{ip_\mu (x^\mu-y^\mu)}\frac{\theta(-p^2-4\mu^2)}{|p^2|} \mathcal{P}_{\mu\nu\rho\sigma} |_{x^\mu=X^\mu(t), y^\mu=X^\mu (t')}. 
  \label{mini_ENtwo}
\end{align}
The derivation of this inequality is presented in Appendix \ref{Ap:Derivation_spectrum_environment}.
The left hand side of Eq.\eqref{mini_ENtwo} gives the following minimum power spectral density,
\begin{align}
    (s^h_{x,\text{env}})^2
    &=
    \frac{1}{m^2 \omega^4}
    \int dt e^{i\omega t}\Delta^\text{min}_{xxxx}(t,0) 
    \notag \\
    &=
    \frac{-2}{27\pi m_p^2}
    \theta(\omega-2\mu)
    \left[
        \frac{\sqrt{\omega^2-4\mu^2}}{5\omega^4}
        \left(
        24\mu^4
        -2\mu^2\omega^2
        +14\omega^4
        \right)
        -3 \omega \,
        \mathrm{Arccoth}\!\left(
        \frac{\omega}{\sqrt{\omega^2-4\mu^2}}
        \right)
    \right].
    \label{mini_ENmodel}
\end{align}
This is independent of the specific functional forms of \eqref{EOp_Noise} and \eqref{EOp_Deco}.
In Fig.~\ref{fig:plot2_OP}, this minimal strain spectrum  $s^h_{x,\text{env}}$ is plotted as a red dot-dashed line, 
while the $D_0$-dependent strain spectrum $S^h_{x,\text{env}}$ given by Eq.~\eqref{Spectrum_Environmental_Oppenheim} is shown as a red solid line.

What this model has in common with the original model plotted in the previous sections is the existence of a minimum fluctuation and the fact that the solid line bends in a way that reflects the trade-off relation. 
On the other hand, there are also notable qualitative differences.
First, the spectrum is a monotonically increasing function of frequency.
Moreover, in contrast to the original  Oppenheim \emph{et al}. model, the low-frequency regime is dominated by $S^N_{x,\text{env}}$ calculated from the noise kernel, whereas the high-frequency regime is governed by $S^D_{x,\text{env}}$ originated from the decoherence kernel (Fig.~\ref{fig:plot2_OP}).
The power spectral density is proportional to $\omega^3$ in the high-frequency regime and $\omega$ in the low-frequency regime.
This behavior reflects that $\Delta^N_{abcd}$ contains the retarded Green's function, which leads to the appearance of a factor scaling as $\sim 1/\omega^{2}$ in its frequency dependence.
In the present analysis, we adopt
$\mu = 10^{-18}\,\mathrm{Hz}$ as the minimal infrared scale, corresponding to the age of the universe.

For comparison, we also plot the power spectral density given by the vacuum fluctuation of a quantized gravitational field predicted in perturbative quantum gravity~\cite{PhysRevD.103.044017}, 
\begin{equation}
  S^h_{x,q}
  =
  \frac{\sqrt{4\pi\omega}}{m_p},
  \label{PSD_QG}
\end{equation}
which is shown as the yellow line in Fig.\ref{fig:plot2_OP}.

As will be discussed in the next section, the magnitude of $\mu$ and that of the resulting spectrum are inversely related.
For the small value of $\mu$, which corresponds to the large strain in \eqref{mini_ENmodel} (red dot-dashed line), the strain predicted in the present model exhibits qualitatively almost the same behavior as that predicted by perturbative quantum gravity (yellow line).
This observation suggests that, depending on the realistic values of the parameters, it may be difficult for experiments to distinguish whether gravity appears effectively classical due to environmental noise or genuinely exhibits quantum behavior. 
In other words, this suggests that the CQ model may mimic perturbative quantum gravity.

\subsection{Short summary for the strain spectra}
\label{sec:shortSUM_for_spectra}

\begin{figure}[ht]
    \centering
    \includegraphics[width=0.95\linewidth]{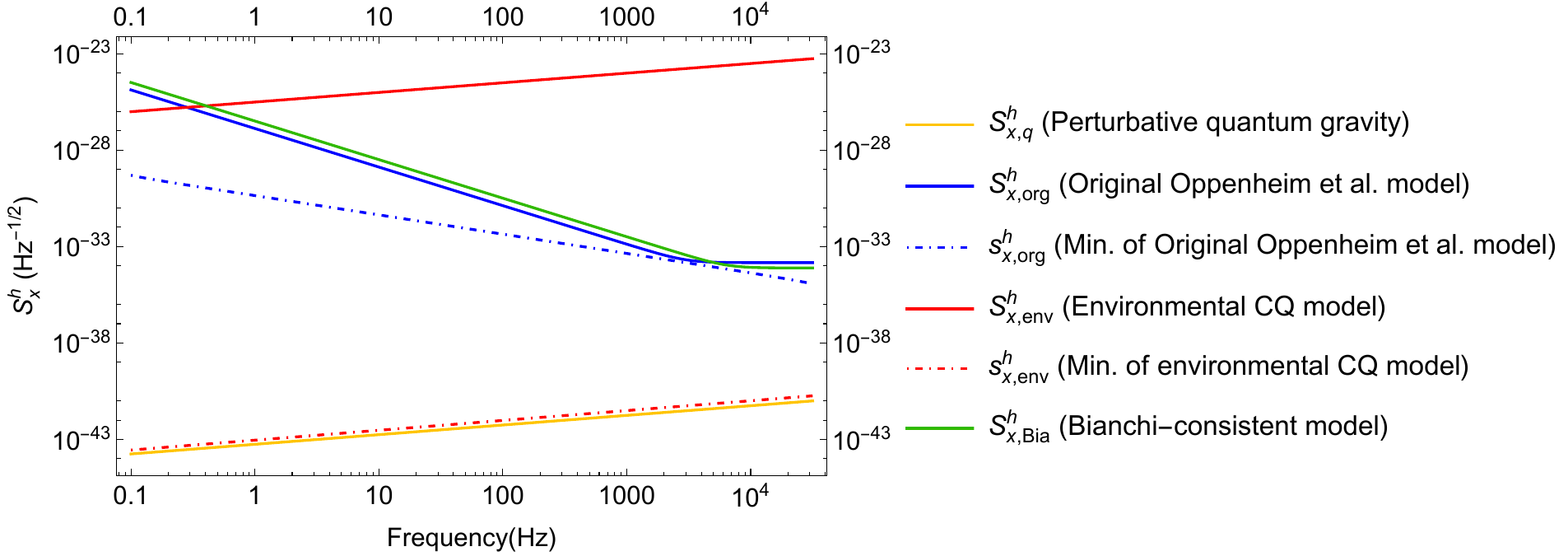}
    \caption{This figure shows the spectra \(S_{x}^h\) for each model plotted together with $D_0 = 10^{-85}\,\mathrm{Hz}^{-4}$.
The blue solid line represents the spectrum of the original Oppenheim \emph{et al}. model, and the blue dot-dashed line indicates its minimum value with the parameter values $\beta = 0.1$, $\epsilon = 10^{-18}\,\mathrm{Hz}$, and $L = 4\,\mathrm{km}$.
The red solid line represents the spectrum of the environmental CQ model, and the red dot-dashed line shows the minimum strain given by Eq.~\eqref{mini_ENmodel} with $\mu = 10^{-18}\,\mathrm{Hz}$.
The green line corresponds to that of the Bianchi-consistent model plotted with the same parameter values as the original Oppenheim \emph{et al}. model, and the yellow line corresponds to the strain predicted from perturbative quantum gravity (vacuum state). 
}
    \label{fig:All_plot}
\end{figure}

Finally, Fig.~\ref{fig:All_plot} shows a simultaneous plot of all the models discussed so far for the purpose of comparison.
The blue solid line represents Eq.~\eqref{Spectrum_Oppenheim_original} for the original Oppenheim \emph{et al}. model, while the blue dot-dashed line corresponds to its minimum, given by Eq.~\eqref{mini_Spectrum_Oppenheim_original}. The green solid line denotes the Bianchi-consistent model in Eq.~\eqref{Spectrum_white_Ein}. The red solid line represents the environmental CQ model in Eq.~\eqref{Spectrum_Environmental_Oppenheim}, and the red dot-dashed line shows its minimum strain, given by Eq.~\eqref{mini_ENmodel}, which is independent of the functional forms of $D_{\mu\nu\rho\sigma}(x,y)$ and $N_{\mu\nu\rho\sigma}(x,y)$. Finally, the yellow solid line corresponds to the strain yielded by perturbative quantum gravity from Ref.~\cite{PhysRevD.103.044017}, as given in Eq.~\eqref{PSD_QG}.

\section{Experimental constraints}
\label{sec:Experimental_constraints}

As seen in the previous section, each model depends on a model-specific parameter $D_0$, and the amplitude of the strain spectrum varies depending on its value. 
In this section, our aim is to place 
expected constraints on this parameter by considering current and future experiments.
As a previous work, Grudka et al. estimated bounds on the parameter $D_0^{\mathrm{\,org}}$ in \cite{grudka2024Renormalisation}, 
with an upper bound derived from interference experiments with large organic molecules~\cite{Gerlich2011,Fein2019} 
and a lower bound obtained from relative acceleration measurements by LISA Pathfinder~\cite{PhysRevLett.120.061101,PhysRevD.110.042004}, 
leading to a constraint $10^{-63} < G_N^2 /D_0^{\mathrm{\,org}} < 10^{-54}$, that is,
\begin{equation}
  10^{-119} < D_0^{\mathrm{\,org}} < 10^{-110}\, \mathrm{Hz}^{-4}.
\label{D0cnstr1}
\end{equation}
To these bounds, we further add constraints derived from the fluctuations of the geodesic deviation calculated in this work.

\begin{figure}[H]
  \centering
  \begin{subfigure}{0.47\linewidth}
    \includegraphics[width=\linewidth]{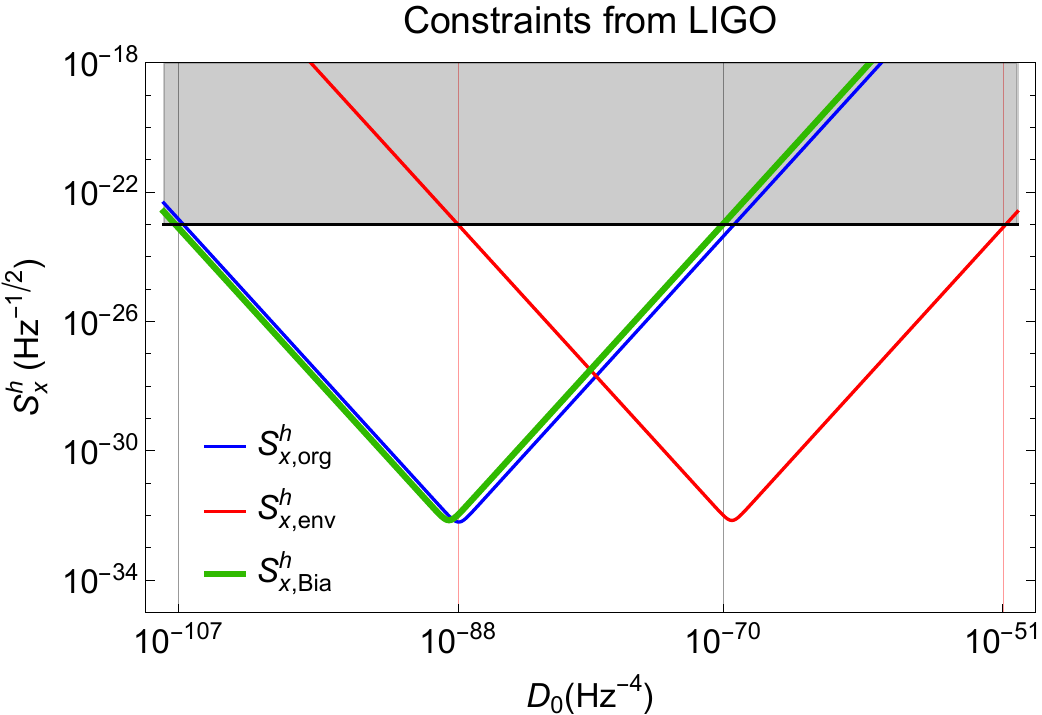}
  \end{subfigure}
  \begin{subfigure}{0.47\linewidth}
\includegraphics[width=\linewidth]{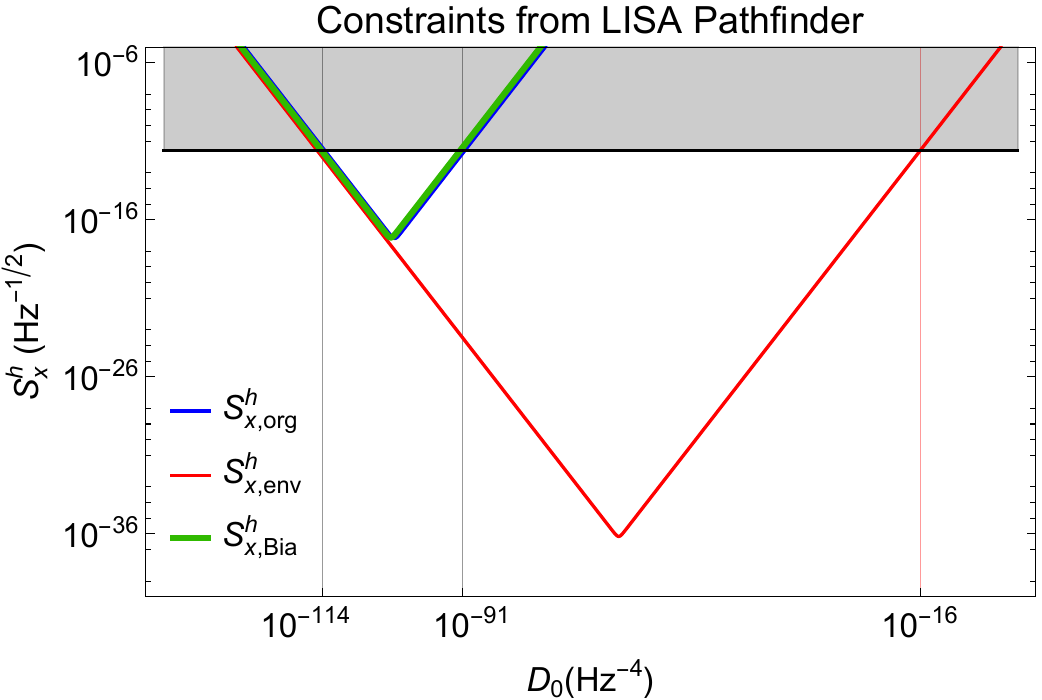}
  \end{subfigure}
  \caption{Constraints on the parameter $D_0$ obtained from the sensitivities of various current gravitational-wave detectors. The spectrum $S^h_{x,\text{org}}$ is plotted in blue with the parameter $\beta = 0.1$ and $\epsilon = 10^{-18}\,\mathrm{Hz}$. 
The environment-induced spectrum $S^h_{x,\text{env}}$ is shown in red and is plotted with $\mu = 10^{-18}\,\mathrm{Hz}$. 
The spectrum $S^h_{x,\text{Bia}}$ is plotted in green using the same parameter as $S^h_{x,\text{org}}$.
The gray shaded region and the black horizontal line indicate the observable regions and the corresponding detection thresholds for each gravitational-wave interferometer. 
In other words, for a given model to remain viable, its predicted spectrum must lie below these regions.
Since mainly $S^h_{x,\text{org}}$ and $S^h_{x,\text{Bia}}$ are very close to each other, the corresponding values of $D_0$ at which each strain spectrum intersects the boundary of the detectable region are approximately indicated by the black vertical lines. Similarly, the red vertical lines indicate the intersection with the environmental CQ model.
Left panel: the constraints from LIGO experiment. The typical sensitivity used here is $10^{-23} \mathrm{Hz}^{-\frac{1}{2}}$ at 100 Hz \cite{LIGO_sens}, and the mean separation $L$ is set to be $4\,\text{km}$. 
Right panel: the constraints from LISA Pathfinder experiment. The sensitivity here is  $10^{-12}\,\mathrm{Hz}^{-\frac{1}{2}}$ at 0.01 Hz {\cite{LISApath_sens}}, and the mean separation $L$ is assumed to be $37.6\,\text{cm}$.}
  \label{fig:Constraint_current}
\end{figure}

Fig.~\ref{fig:Constraint_current} shows the spectra for the models, $S^h_{x,\text{org}}$, $S^h_{x,\text{Bia}}$ and $S^h_{x,\text{env}}$ given by \eqref{Spectrum_Oppenheim_original},\eqref{Spectrum_white_Ein} and \eqref{Spectrum_Environmental_Oppenheim}, respectively. 
For all plots, all parameters except for $D_0$ are fixed.
The left panel of Fig.~\ref{fig:Constraint_current} shows the constraints from LIGO experiment. 
The mean separation length $L$ is roughly taken to be $4 \, \text{km}$. 
The strain sensitivity of LIGO is roughly estimated as $10^{-23}\,\text{Hz}^{-1/2}$ around the frequency 
$\omega \sim 100\,\mathrm{Hz}$ \cite{LIGO_sens}. 
If the strain spectrum calculated for each of the three CQ models exceeds $10^{-23}\,\text{Hz}^{-1/2}$, 
the model with such a strain is negative because it would be not observed in LIGO\footnote{Strictly speaking, to get actual observational constraints, we should analyze optical readouts predicted in CQ models assuming LIGO-type experiments. Here, we simply put expected constraints from strain sensitivities.}.  
Requiring that each strain of the models is below the threshold sensitivity, we get allowed parameter ranges. 
For example, the allowed range of $D^\text{org}_0$ for the original Oppenheim \emph{et al}. model, shown in blue line, is estimated to be 
\begin{equation}
  10^{-107} < D_0^{\mathrm{\,org}} < 10^{-69}\, \mathrm{Hz}^{-4}.
  \label{D0cnstr2}
\end{equation}
This constraint is obtained by assuming $\beta=0.1$. As discussed in Sec. \ref{section:Oppenheim original model}, we have the singular behavior of the $\beta$ dependence in the original Oppenheim \emph{et al.} model. 
Particularly, when $\beta = 1/3$, the noise contribution $S^\text{N}_{x,\text{org}}$ vanishes, and the lower bound on $D_0^{\mathrm{\,org}}$ disappears. We then can enlarge the allowed parameter range of $D^\text{org}_0$.
The lower bound in Eq.~\eqref{D0cnstr2} was also discussed in Ref.~\cite{oppenheim2025diffusionstochastickleingordonequation}. There, it was given as $N_0^{\mathrm{\,org}}=(4\pi G_N)^2/D_0^{\mathrm{\,org}}\lesssim10^{-65}$ by analyzing field dynamics with stochastic noise\footnote{Their bound was inferred from an analysis of a scalar field rather than the gravitational field. However, the calculation of the two-point correlation function for their scalar field
is almost equivalent to that for our metric perturbations $h_{\mu\nu}$.}, and this estimate is close to the result obtained in our analysis.
On the other hand, we have the upper bound in Eq.~\eqref{D0cnstr2} by considering the contribution of decoherence to the fluctuations of geodesic deviation.

For the strains of the Bianchi-consistent model in Sec.\ref{sec:CQwithEin} and the environmental CQ model in Sec.\ref{sec:CQwithEnv}, which are shown in green and red lines in Fig.~\ref{fig:Constraint_current}, respectively, one obtains the constraints
\begin{equation}
   10^{-107} < D_0^{\mathrm{\,Bia}} < 10^{-70}\, \mathrm{Hz}^{-4} ,\quad  10^{-88} < D_0^{\mathrm{\,env}} < 10^{-51}\, \mathrm{Hz}^{-4}.
\end{equation}
The constraints on $D_0^{\mathrm{\,Bia}}$ and $D_0^{\mathrm{\,env}}$ are, as in the case of $D_0^{\mathrm{\,org}}$, such that the lower bound arises from the contribution of noise, while the upper bound arises from the contribution of decoherence.
To get the constraint on $D^\text{env}_0$, we set the energy scale of the environmental degrees of freedom to be $\mu=10^{-18}\text{Hz}$. The energy scale plays a crucial role 
in determining the strength of the constraints.
Due to the effect of the step function appearing in Eq.~\eqref{Spectrum_Environmental_Oppenheim}, decreasing $\mu$ increases the number of contributing 
$\omega$ modes to gravitational fluctuations, thereby enhancing the noise and leading to more stringent constraints. 
Conversely, increasing $\mu$ reduces the number of contributing modes and relaxes the observational bounds.

For comparison, the sensitivity of another operating experiment, LISA Pathfinder, is also shown in the right panel of Fig.~\ref{fig:Constraint_current}.
Here, the mean separation length $L$ was estimated to be about $37.6\,\text{cm}$.
At present, however, the constraint derived from LIGO is more stringent. 
Therefore, in the following discussion we mainly rely on the results shown in the left panel of Fig.~\ref{fig:Constraint_current}.

\begin{figure}[H]
  \centering
  \begin{subfigure}{0.45\linewidth}
    \includegraphics[width=\linewidth]{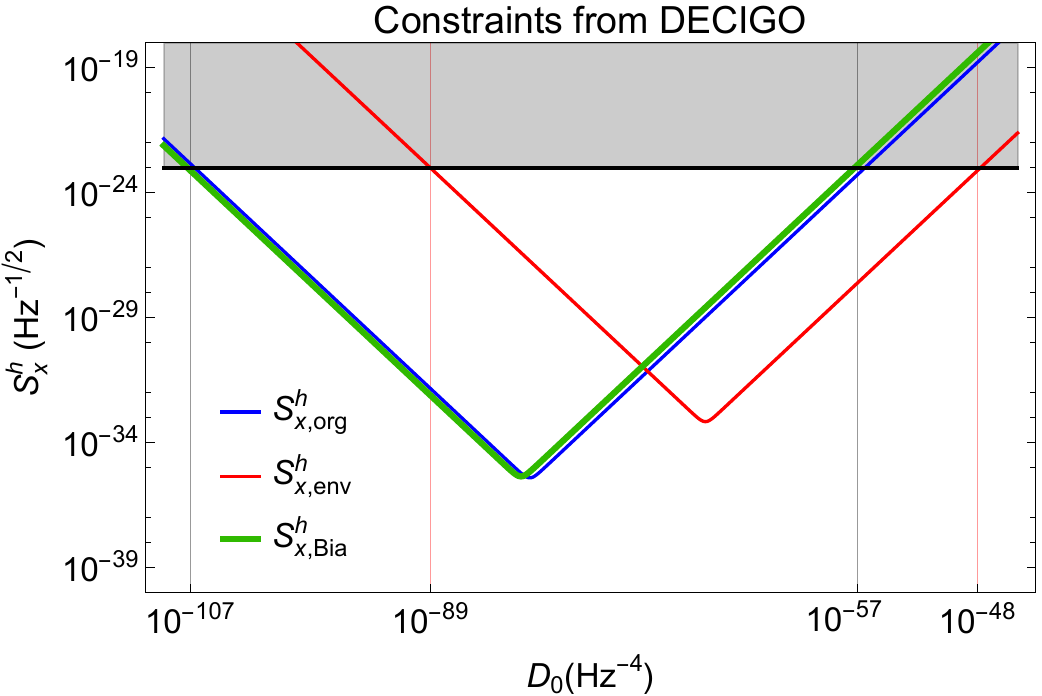}
  \end{subfigure}
  \begin{subfigure}{0.45\linewidth}
    \includegraphics[width=\linewidth]{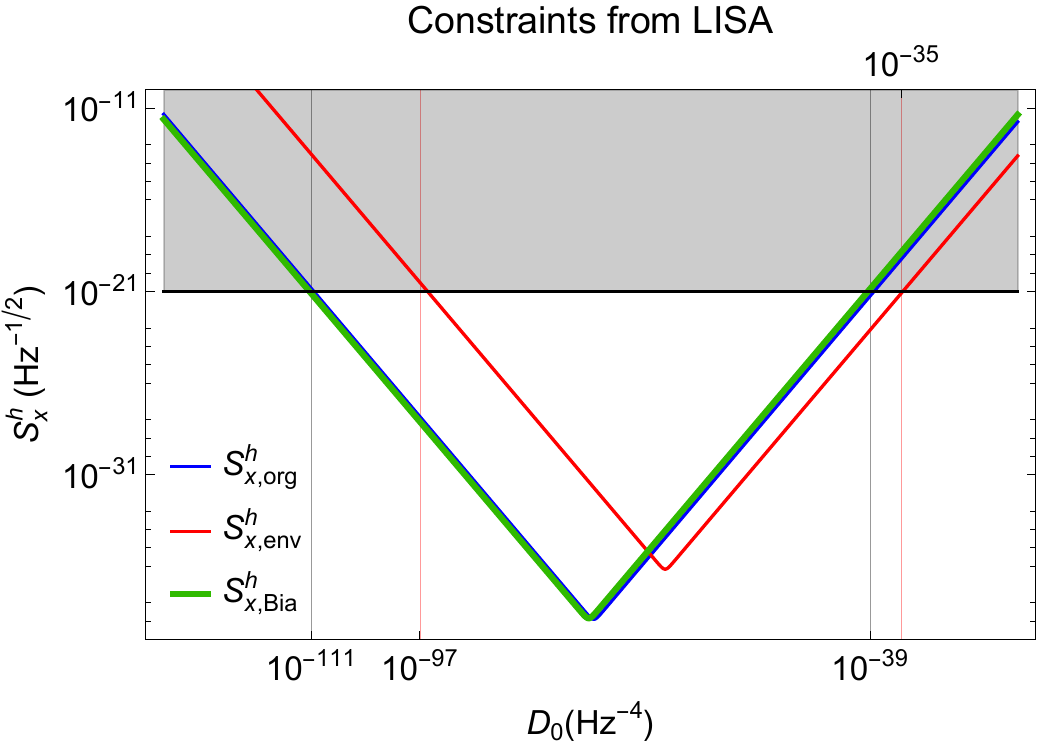}
  \end{subfigure}
  
  \vspace{2.5em}
  
  \begin{subfigure}{0.45\linewidth}
    \includegraphics[width=\linewidth]{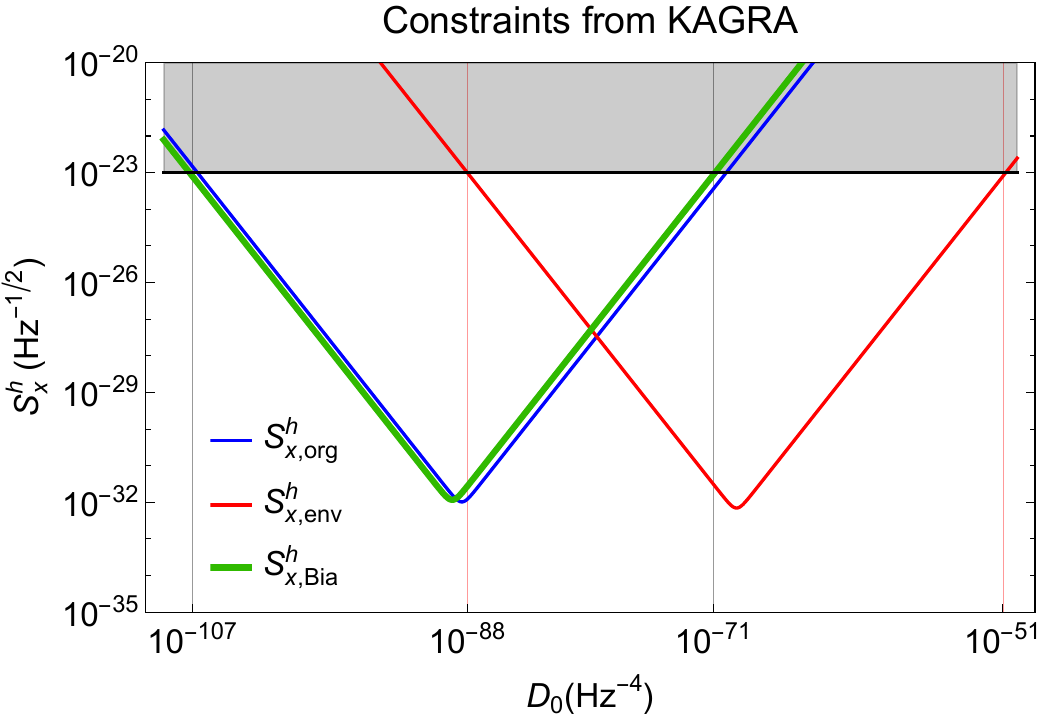}
  \end{subfigure}
  \begin{subfigure}{0.45\linewidth}
    \includegraphics[width=\linewidth]{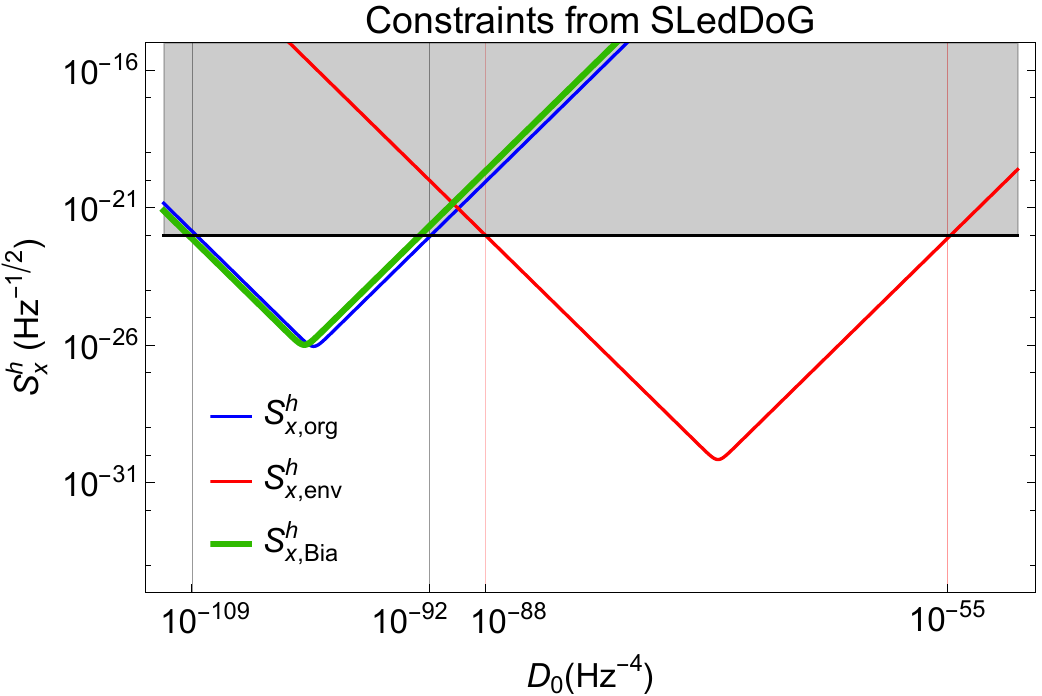}
  \end{subfigure}

  \caption{Constraints on the parameter $D_0$ obtained from the sensitivities of various future gravitational-wave detectors. 
  The spectrum $S^h_{x,\text{org}}$ is plotted in blue with the parameter $\beta = 0.1$ and $\epsilon = 10^{-18}\,\mathrm{Hz}$. 
The environment-induced spectrum $S^h_{x,\text{env}}$ is shown in red and is plotted with $\mu = 10^{-18}\,\mathrm{Hz}$. 
The spectrum $S^h_{x,\text{Bia}}$ is plotted in green using the same parameter as $S^h_{x,\text{org}}$.
The gray shaded region and the black horizontal line indicate the observable regions and the corresponding detection thresholds for each gravitational-wave interferometer. 
In other words, for a given model to remain viable, its predicted spectrum must lie below these regions.
Since mainly $S^h_{x,\text{org}}$ and $S^h_{x,\text{Bia}}$ are very close to each other, the corresponding values of $D_0$ at which each strain spectrum intersects the boundary of the detectable region are approximately indicated by the black vertical lines. Similarly, the red vertical lines indicate the intersection with the environmental CQ model.
The sensitivities and the mean lengths used for the left upper (DECIGO), right upper (LISA), left lower (KAGRA) and right lower (SLedDoG) panels are $10^{-23}\,\mathrm{Hz}^{-\frac{1}{2}}$ at 10 Hz and $L=1000\,\mathrm{km}$ \cite{DECIGO_sens}, $10^{-21}\,\mathrm{Hz}^{-\frac{1}{2}}$ at 0.01 Hz and $L=5\times10^{6}\,\mathrm{km}$ \cite{LISA_sens}, 
$10^{-23}\,\mathrm{Hz}^{-\frac{1}{2}}$ at 100 Hz and $L=3\,\mathrm{km}$ \cite{KAGRA_sens}, and $10^{-22}\,\mathrm{Hz}^{-\frac{1}{2}}$ at $10^4$ Hz and $L=8.6\,\mathrm{cm}$ \cite{SLedDoG}, respectively.
}
  \label{fig:Constraint_future}
\end{figure}

When the constraints of \eqref{D0cnstr1} and \eqref{D0cnstr2} are naively combined, 
one finds that the original Oppenheim \emph{et al}. model is observationally excluded. 
Also, if the bound of $D_0^\text{ori}$, Eq.\eqref{D0cnstr1}, can be applied to $D^\text{Bia}_0$ and $D^\text{env}_0$, the models with the strains $S^h_{x,\text{Bia}}$ and $S^h_{x,\text{env}}$ would be ruled out. 
However, the minimal strain $s^h_{x,\text{env}}$ given in Sec.\ref{sec:CQwithEnv}, which can be achieved by tuning the noise kernel and the decoherence kernel, is not excluded since it is extremely small. 
For example, as observed in Fig. \ref{fig:All_plot}, the minimal strain $s^h_{x,\text{env}}$ is about $10^{-42}\,\text{Hz}^{-1/2}$ around $\omega \sim 100\,\text{Hz}$, which is much smaller than the LIGO sensitivity and is not detectable in the LIGO experiment.
Furthermore, the constraints on the parameters from gravitational-wave experiments planned in the near future 
are summarized in Fig.~\ref{fig:Constraint_future}. 
Since the mean separation $L$ between two massive objects and the frequency band of highest sensitivity differ among experiments, which models can be effectively constrained depends strongly on the characteristics of each experiment.

In addition, the stochastic backgrounds predicted by the CQ models considered here may be distinguishable from conventional astrophysical stochastic gravitational-wave backgrounds through their characteristic spectral shapes. For example, in the frequency range relevant to ground-based detectors such as LIGO, unresolved compact binary inspirals are expected to produce an energy-density spectrum approximately given by $\Omega_{\rm GW}(\omega)\propto \omega^{2/3}$ in the range $50$--$100,{\rm Hz}$ \cite{phinney2001,Zhu2013}. Using the relation
$(S^h(\omega))^2=\frac{3H_0^2}{2\pi^2}\omega^{-3}\Omega_{\rm GW}(\omega)$,
the corresponding strain spectrum behaves as $S^h(\omega)\propto \omega^{-7/6}$ \cite{MAGGIORE2000283}. By contrast, in the lower-bound regime derived in the present work, the spectrum is dominated by the noise contribution. As a result, the strain spectra of $S_{\rm org}$ and $S_{\rm Bia}$ are approximately frequency independent, whereas the environmental CQ model predicts a strain spectrum proportional to $\omega$. Therefore, the CQ-induced stochastic backgrounds considered here may be distinguishable from conventional astrophysical backgrounds through their different frequency dependences.

\section{Conclusion and Discussion}

In this paper, we analyzed the fluctuations of geodesic deviation between quantum objects in a classical gravitational field within the framework of the relativistic semiclassical gravity model proposed by Oppenheim \emph{et al}.
We found that the noise introduced in the original Oppenheim \emph{et al}. model may not strictly satisfy the Bianchi identities. Motivated by this observation, we constructed a modified Oppenheim \emph{et al}. model that is manifestly consistent with Bianchi identities and performed the same analysis. Nevertheless, the computed spectrum showed no significant qualitative differences.

Our analysis suggests that models based on the simple white noise kernel adopted in previous studies \cite{PhysRevX.13.041040,oppenheim2025covariant,grudka2024Renormalisation,Oppenheim2023_Nature} can be readily tested by current gravitational-wave detectors, such as LIGO, whether such models can be ruled out or remain effectively viable, if we combine our constraints to another constraint in Ref.~\cite{grudka2024Renormalisation}.
As pointed out in Chapters \ref{section:Oppenheim original model} and \ref{sec:Experimental_constraints}, for $\tfrac{1}{4} \le \beta \le \tfrac{1}{3}$ neither a valid spectrum nor a constraint on $D_0^{\mathrm{\,org}}$ can be obtained. However, in the Bianchi-consistent model that we proposed in Sec.\ref{sec:CQwithEin}, the spectrum is obtained independently of $\beta$. Therefore, the constraint on $D_0^{\mathrm{\,org}}$ derived here is still expected to remain valid.
If the constraint of Ref.~\cite{grudka2024Renormalisation} is naively applied to $D_0^{\mathrm{\,env}}$, our environmental CQ model might also be ruled out; however, this point remains open to discussion.

Moreover, from a theoretical perspective, we demonstrated that assuming such scale-free noise inevitably leads to a divergence of observables after sufficiently long time evolution. This divergence may originate from the approximation in which dissipative effects due to gravitational wave radiation from the motion of the deviation are neglected. It therefore remains an important open problem to investigate whether similar divergences persist when dissipation is properly taken into account.

Furthermore, we considered the origin of the stochastic fluctuations intrinsic to the classical gravitational field, which was not clearly specified in the original Oppenheim \emph{et al}. model, and we constructed a model in which the gravitational field itself is assumed to be quantum mechanical and subject to environmental fluctuations through interactions with other quantum fields, and carried out the same analysis. And we found that the frequency dependence of the spectrum, as well as the regimes dominated by noise and decoherence, exhibit behavior opposite to that of the original Oppenheim \emph{et al}. model.
However, in both cases there exists a minimum fluctuation that is independent of the parameter $D_{0}$.

In this environmental CQ model, colored noise arises naturally as a consequence of assuming that the gravitational field is fundamentally quantum and coupled to environmental degrees of freedom. However, even if gravity were fundamentally classical, the possibility that effective colored noise could emerge cannot be excluded. Therefore, even if this model were to be experimentally supported, one should be cautious in interpreting the result as direct evidence for the intrinsic quantum nature of gravity, as opposed to an effectively classical description.

On the other hand, white noise is expected to arise naturally only when the gravitational field is fundamentally classical. In this sense, the exclusion of white-noise-type models can be regarded as an important step toward probing the quantum nature of gravity. In addition, the environmental CQ model does not require artificial ingredients such as UV cutoff $L$ that were necessary in the original Oppenheim \emph{et al}. model, and thus provides a theoretically more consistent framework among models in which gravity behaves effectively classically.
However, regarding the IR cutoff $\epsilon$ required in the original Oppenheim \textit{et al.} model, it may be that in this model it is merely replaced by the environmental degree-of-freedom parameter $\mu$. As $\mu$ increases, the number of frequency modes contributing to the noise in the gravitational field decreases, and consequently the fluctuation of the geodesic deviation becomes smaller. From the viewpoint of constraining the model, this works in an unfavorable direction. Therefore, if the energy scale of the assumed environment is large, the constraints become weaker than those indicated by the plots shown in Figs.~\ref{fig:Constraint_current} and \ref{fig:Constraint_future}.

In addition, the stochastic backgrounds considered in this work are expected to be approximately Gaussian, since the stochastic source $\chi_{\mu\nu}$ is Gaussian by construction. In principle, higher-order correlators such as the bispectrum or trispectrum could provide additional information for distinguishing such backgrounds from astrophysical stochastic backgrounds, which may exhibit non-Gaussian features arising from unresolved discrete sources. However, a more quantitative investigation of this possibility would require the construction and analysis of CQ models with intrinsically non-Gaussian stochastic sources, which we leave for future work.

Finally, we found that the behavior of the minimum strain spectrum, which is independent of the specific functional forms of noise and decoherence in our environmental CQ model is close to that obtained in perturbative quantum gravity. 
This suggests that the CQ model may be experimentally difficult to distinguish from perturbative quantum gravity, or that it can effectively mimic its predictions.
And it also suggests that even within an effective classical-gravity description, the question of whether quantum entanglement can be generated remains highly nontrivial and constitutes an important direction for future research.

\acknowledgments{
We would like to express sincere gratitude to Daniel Carney, Laurent Freidel, Youka Kaku, Manthos Karydas, Adrian Kent, Isaac Layton, Giacomo Marocco, Amaury Micheli, Kota Numajiri, Emanuele Panella, Yuta Uenaga, and Yuko Urakawa for valuable discussions and insightful comments.
In particular, we are deeply grateful to Yutaka Shikano for fruitful discussions from an experimental perspective throughout the course of this work.
A.M. was supported by JSPS
KAKENHI (Grants No. JP23K13103 and No. JP23H01175).
}

\appendix

\section{Derivation of the action for geodesic deviation}
\label{Ap:Action_for_geodesic_devi}

In this appendix, we derive the action for a geodesic deviation, $S_\text{dev}=\frac{m}{2}\int dt \left( \delta_{ab} \frac{d\xi^{a}}{dt}\frac{d\xi^{b}}{dt}-R^{(1)}_{0a0b} \xi^a \xi^b \right)$, given as the first part of Eq.\eqref{effective_action}.
The action for masses $M$ and $m$, which gives geodesics of each mass, is 
\begin{align}
    S
    =-M\int d\lambda \sqrt{-g_{\mu\nu}(x)\frac{dx^{\mu}}{d\lambda} \frac{dx^{\nu}}{d\lambda}}
    -
    m\int d\lambda \sqrt{-g_{\mu\nu}(y)\frac{dy^{\mu}}{d\lambda} \frac{dy^{\nu}}{d\lambda}}. 
\end{align}
Here we assume that $x^\mu$ follow the geodesic equations, and let $\tau$ denote the proper time along the geodesic $x^\mu$. 
We consider a small spacetime separation between $M$ and $m$ given by
\begin{equation}
    y^{\mu} = x^{\mu} + \xi^{\mu},
\end{equation}
and expand the action associated with $y^{\mu}$ up to second order in $\xi^{\mu}$.
Introducing $u^{\mu}=\dot{x}^{\mu}=dx^{\mu}/d\tau$ and $\dot{\xi}^{\mu}=d\xi^{\mu}/d\tau$, we obtain
\begin{align}
    -g_{\mu\nu}(x+\xi) (u^{\mu}+\dot\xi^{\mu}) (u^{\nu}+\dot\xi^{\nu})
    &=
    \left[ g_{\mu\nu}(x) + \xi^{\alpha} \partial_{\alpha}g_{\mu\nu}(x) + \frac{1}{2}\xi^{\alpha}\xi^{\beta} \partial_{\alpha}\partial_{\beta} g_{\mu\nu}(x) + \cdots \right](u^{\mu}+\dot\xi^{\mu}) (u^{\nu}+\dot\xi^{\nu}) \notag\\
    &\approx -1+2g_{\mu\nu} u^{\mu}\dot\xi^{\nu}
    +\xi^{\alpha}\partial_{\alpha} g_{\mu\nu} u^{\mu} u^{\nu} + g_{\mu\nu} \dot\xi^{\mu} \dot\xi^{\nu}+2\xi^{\alpha}\partial_{\alpha} g_{\mu\nu} u^{\mu} \dot\xi^{\nu}
    +\frac{1}{2} \xi^{\alpha} \xi^{\beta} \partial_{\alpha} \partial_{\beta} g_{\mu\nu} u^{\mu} u^{\nu}.
\end{align}
Expanding the action for $y^{\mu}$, we get the action for the deviation $\xi^\mu$. 
Explicitly, the expansion is evaluated as 
\begin{align}
    S_\text{dev}
    &=-m\int d\tau\sqrt{-g_{\mu\nu}(y)\frac{dy^{\mu}}{d\tau} \frac{dy^{\nu}}{d\tau}} \notag\\
    &\approx
    -m \int d\tau \Bigg[ 1-g_{\mu\nu} u^{\mu}\dot\xi^{\nu}
    -\frac{1}{2}\xi^{\alpha}\partial_{\alpha} g_{\mu\nu} u^{\mu} u^{\nu} - \frac{1}{2}g_{\mu\nu} \dot\xi^{\mu} \dot\xi^{\nu}
    -\xi^{\alpha}\partial_{\alpha} g_{\mu\nu} u^{\mu} \dot\xi^{\nu}
    -\frac{1}{4} \xi^{\alpha} \xi^{\beta} \partial_{\alpha} \partial_{\beta} g_{\mu\nu} u^{\mu} u^{\nu} \notag\\
    &\qquad
    -\frac{1}{8} \left( 2g_{\mu\nu} u^{\mu}\dot\xi^{\nu}
    + \xi^{\alpha}\partial_{\alpha} g_{\mu\nu} u^{\mu} u^{\nu} \right)^2
    \Bigg].
    \label{S_y_1}
\end{align}
This can be rewritten in a manifestly covariant way.
Introducing the covariant derivative along with the geodesic $x^\mu$ with the Levi-Civita connection $\Gamma^\mu_{\rho \sigma}$, 
\begin{equation}
  \frac{D \xi^{\mu}}{D \tau}
  = \dot\xi^{\mu} +\Gamma^{\mu}_{\rho\sigma} u^{\rho} \xi^{\sigma},
\end{equation}
 and using the metricity $\nabla_{\alpha} g_{\mu\nu}=\partial_{\alpha} g_{\mu\nu}
  -\Gamma^{\beta}_{\mu \alpha} g_{\beta\nu} - \Gamma^{\beta}_{\alpha \nu} g_{\mu\beta}=0$, 
we find that the first-order terms in $\xi^\mu$ is written as
\begin{align}
  g_{\mu\nu} u^{\mu}\dot\xi^{\nu} + \frac{1}{2} \xi^{\alpha}\partial_{\alpha} g_{\mu\nu} u^{\mu} u^{\nu}
  =\frac{D }{D \tau}(u_{\nu}\xi^{\nu}),
  \label{eq:xiu}
\end{align}
where note that $u^\mu$ follows the equations $D u^\mu /D\tau =0$.
Furthermore, the second-order terms in $\xi^\mu$ have the following expression,
\begin{align}
  &- \frac{1}{2}g_{\mu\nu} \dot\xi^{\mu} \dot\xi^{\nu}
    -\xi^{\alpha}\partial_{\alpha} g_{\mu\nu} u^{\mu} \dot\xi^{\nu}
    -\frac{1}{4} \xi^{\alpha} \xi^{\beta} \partial_{\alpha} \partial_{\beta} g_{\mu\nu} u^{\mu} u^{\nu} -\frac{1}{8} \left( 2g_{\mu\nu} u^{\mu}\dot\xi^{\nu}
    + \xi^{\alpha}\partial_{\alpha} g_{\mu\nu} u^{\mu} u^{\nu} \right)^2\notag \\
&\quad =- \frac{1}{2}g_{\mu\nu} \dot{\xi}^\mu \dot{\xi}^\nu
    -\xi^{\alpha}(\Gamma^{\beta}_{\mu \alpha} g_{\beta\nu} + \Gamma^{\beta}_{\alpha \nu} g_{\mu\beta})u^{\mu} \dot{\xi}^\nu -\frac{1}{4} \xi^{\alpha} \xi^{\beta} \partial_{\alpha}(\Gamma^{\rho}_{\mu \beta} g_{\rho\nu} +\Gamma^{\rho}_{\beta \nu} g_{\mu\rho} )u^{\mu} u^{\nu} -\frac{1}{2} \left( \frac{D}{D\tau} (u^\mu \xi_\mu)\right)^2\notag \\  
 &\quad = -\frac{1}{2}g_{\mu\nu} \Big(\frac{D\xi^\mu}{D\tau}-\Gamma^\mu_{\rho \sigma}u^\rho \xi^\sigma \Big) \Big(\frac{D\xi^\nu}{D\tau}-\Gamma^\nu_{\alpha \beta}u^\alpha \xi^\beta \Big)
 -\xi^{\alpha}u^\mu \Gamma^{\beta}_{\mu \alpha} g_{\beta\nu}\Big(\frac{D\xi^\nu}{D\tau}-\Gamma^\nu_{\rho \sigma} u^\rho \xi^\sigma \Big) 
 -\xi^\alpha \Gamma^{\beta}_{\alpha \nu} g_{\mu\beta}u^{\mu} \dot{\xi}^\nu 
 -\frac{1}{2} \left( \frac{D}{D\tau} (u^\mu \xi_\mu)\right)^2
 \notag\\
&\qquad -\frac{1}{4} \xi^{\alpha} \xi^{\beta} \partial_{\alpha}(\Gamma^{\rho}_{\mu \beta} g_{\rho\nu} +\Gamma^{\rho}_{\beta \nu} g_{\mu\rho} )u^{\mu} u^{\nu} \notag \\  
&\quad = -\frac{1}{2}(g_{\mu\nu}+u_\mu u_\nu) \frac{D\xi^\mu}{D\tau} \frac{D\xi^\nu}{D\tau}+\frac{1}{2}g_{\mu\nu}\Gamma^\mu_{\rho \sigma}u^\rho \xi^\sigma \Gamma^\nu_{\alpha \beta}u^\alpha \xi^\beta
-\frac{1}{2}  \Gamma^{\beta}_{\alpha \nu} g_{\mu\beta}u^{\mu} \frac{d}{d\tau}(\xi^\alpha \xi^\nu) 
-\frac{1}{4} \xi^{\alpha} \xi^{\beta} \partial_{\alpha}(\Gamma^{\rho}_{\mu \beta} g_{\rho\nu} +\Gamma^{\rho}_{\beta \nu} g_{\mu\rho} )u^{\mu} u^{\nu} \notag \\  
&\quad = -\frac{1}{2}(g_{\mu\nu}+u_\mu u_\nu) \frac{D\xi^\mu}{D\tau} \frac{D\xi^\nu}{D\tau}+\frac{1}{2}g_{\mu\nu}\Gamma^\mu_{\rho \sigma}u^\rho \xi^\sigma \Gamma^\nu_{\alpha \beta}u^\alpha \xi^\beta
-\frac{1}{2}  \frac{d}{d\tau} \Big(\Gamma^{\beta}_{\alpha \nu} g_{\mu\beta}u^{\mu} \xi^\alpha \xi^\nu \Big) 
+\frac{1}{2} \frac{d}{d\tau} \Big(\Gamma^{\beta}_{\alpha \nu} g_{\mu\beta}u^{\mu} \Big) \xi^\alpha \xi^\nu \notag \\
& \qquad -\frac{1}{4} \xi^{\alpha} \xi^{\beta} \partial_{\alpha}(\Gamma^{\rho}_{\mu \beta} g_{\rho\nu} +\Gamma^{\rho}_{\beta \nu} g_{\mu\rho} )u^{\mu} u^{\nu}.  
\label{eq:xixiuu}
\end{align}
Substituting Eqs.\eqref{eq:xiu} and \eqref{eq:xixiuu}  into \eqref{S_y_1}, we get
\begin{align}
  S_\text{dev}
&=-m\int d\tau \Big[ -\frac{1}{2}(g_{\mu\nu}+u_\mu u_\nu) \frac{D\xi^\mu}{D\tau} \frac{D\xi^\nu}{D\tau}
+\frac{1}{2}g_{\mu\nu}\Gamma^\mu_{\rho \sigma}u^\rho \xi^\sigma \Gamma^\nu_{\alpha \beta}u^\alpha \xi^\beta
+\frac{1}{2} \frac{d}{d\tau} \Big(\Gamma^{\beta}_{\alpha \nu} g_{\mu\beta}u^{\mu} \Big) \xi^\alpha \xi^\nu \notag\\
&\quad -\frac{1}{4} \xi^{\alpha} \xi^{\beta} \partial_{\alpha}(\Gamma^{\rho}_{\mu \beta} g_{\rho\nu} +\Gamma^{\rho}_{\beta \nu} g_{\mu\rho} )u^{\mu} u^{\nu}\Big],
  \label{S_y_2}
\end{align}
where we ignored the surface term $1- D (u_{\nu}\xi^{\nu})/D \tau -\frac{1}{2}  d \Big(\Gamma^{\beta}_{\alpha \nu} g_{\mu\beta}u^{\mu} \xi^\alpha \xi^\nu \Big)/d\tau $ that does not contribute to the equations of motion under appropriate boundary conditions.
Moreover, the geodesic equations $D u^\mu/D\tau=0$ and the Riemann tensor
\begin{align}
 R_{\mu \alpha \nu}{}^{\beta}=\partial_\mu \Gamma^\beta_{\alpha \nu}- \partial_\alpha \Gamma^\beta_{\mu \nu}+\Gamma^\lambda_{\alpha \nu} \Gamma^\beta_{\lambda \mu}-\Gamma^\lambda_{\mu \nu } \Gamma^\beta_{\lambda \alpha },
\end{align}
allow us to find that the terms other than the first kinetic term in the above action $S_y$ take the simple form,
\begin{align}
 \frac{1}{2}g_{\mu\nu}\Gamma^\mu_{\rho \sigma}u^\rho \xi^\sigma \Gamma^\nu_{\alpha \beta}u^\alpha \xi^\beta
+\frac{1}{2} \frac{d}{d\tau} \Big(\Gamma^{\beta}_{\alpha \nu} g_{\mu\beta}u^{\mu} \Big) \xi^\alpha \xi^\nu -\frac{1}{4} \xi^{\alpha} \xi^{\beta} \partial_{\alpha}(\Gamma^{\rho}_{\mu \beta} g_{\rho\nu} +\Gamma^{\rho}_{\beta \nu} g_{\mu\rho} )u^{\mu} u^{\nu}=\frac{1}{2} R_{\mu \alpha \nu \beta} u^\mu u^\nu \xi^\alpha \xi^\beta.
  \label{eq:Ruuxixi}
\end{align}
Substituting this relation into \eqref{S_y_2}, we obtain
\begin{align}
  S_\text{dev}
  &=\frac{m}{2} \int d\tau
  \left[
  (g_{\mu\nu} +u^{\mu}u^{\nu})
  \frac{D \xi^{\mu}}{D \tau}
  \frac{D \xi^{\nu}}{D \tau}
  - R_{\mu\alpha\nu\beta}
  u^{\mu}u^{\nu}
  \xi^{\alpha}
  \xi^{\beta}
  \right].
  \label{S_y_4}
\end{align}
Here, we prepare the Fermi normal coordinate by using the tetrad $\{u^\mu, e^\mu_a \}_{a=x,y,z}$ that satisfies
\begin{equation}
 g_{\mu \nu} u^\mu e^\nu_a=0, \quad g_{\mu \nu} e^\mu_a e^\nu_b=\delta_{ab}, \quad \frac{D e^\mu_a}{D\tau}=0,
\end{equation}
along the geodesic $x^\mu$, where 
$\delta_{ab}$ is the Kronecker delta. 
Then the vector $\xi^\mu$ can be decomposed as 
\begin{equation}
\xi^\mu=u^\mu \xi +e^\mu_a \xi^a,
\end{equation}
and substituting this into the above action \eqref{S_y_4}, we get 
\begin{align}
  S_\text{dev}
  &=\frac{m}{2} \int d\tau
  \left[
 \delta_{ab}
  \frac{d \xi^a}{d\tau}
  \frac{d \xi^b}{d \tau}
  - R_{\mu a \nu b}
  u^{\mu}u^{\nu}
  \xi^a
  \xi^b
  \right].
  \label{S_xi}
\end{align}
This is the action of the deviation $\xi^a$ coupled with the spacetime curvature.

Next we introduce a metric perturbation
\begin{equation}
  g_{\mu\nu}(x^{\alpha})=\eta_{\mu\nu}+h_{\mu\nu}(x^{\alpha}),
\end{equation}
and denote a global inertial time, the spacetime position of mass $M$, and its four-velocity in the limit $h_{\mu \nu}\to0$ by $t$, $X^{\mu}$, and $U^{\mu}=dX^\mu/dt$, respectively. They satisfy
$
\eta_{\mu\nu}U^{\mu}U^{\nu}=-1$, $dU^{\mu}/dt=0$. Expanding the action to first order in $h_{\mu\nu}$,  we obtain
\begin{align}
S_\text{dev}
  &\approx
  \frac{m}{2} \int dt \left[
  \delta_{ab} \frac{d\xi^a}{dt} \frac{d\xi^b}{dt}
  - R^{(1)}_{\mu a \nu b} U^{\mu}U^{\nu}\xi^a\xi^b
  \right],
\end{align}
where we assumed that $x^\mu$ and $\xi^a$ behave sufficiently nonrelativistically and neglected $O(h_{\mu \nu} \frac{d\xi^a}{dt}\frac{d\xi^b}{dt})$. Here, $R^{(1)}_{\mu a \nu b}$ is the Riemann tensor up to first order in $h_{\mu \nu}$ given as 

\begin{align}
R^{(1)}_{\mu a \nu b}=\frac{1}{2}[ \partial_a\partial_\nu  h_{\mu b}-\partial_b \partial_a h_{\mu \nu}-\partial_\mu \partial_\nu h_{a b}+\partial_b \partial_\mu h_{a \nu}]. 
\end{align}
When mass $M$ is at rest, $X^\mu (t)=[t,0,0,0]$ and $ U^{\mu}=(1,0,0,0)$, the action further is simplified to 
\begin{equation}
  S_\text{dev}
  =\frac{m}{2}\int dt \left[
  \delta_{ab}\frac{d\xi^a}{dt}\frac{d\xi^b}{dt}
  -R^{(1)}_{0a0b}\xi^a\xi^b
  \right],
\end{equation}
which coincides with the first term of Eq.~\eqref{effective_action}.

\section{Derivation of Langevin equation}
\label{Derivation of Langevan equation}

In this section, starting from the CQ path-integral formulation \eqref{CQ_path_integral}, 
we integrate out the classical gravitational field getting the Feynman--Vernon influence functional, 
and derive a Langevin equation for the geodesic deviation.
We first redefine the geodesic deviation $\xi^a$ in terms of the mean separation $L^a$ of the geodesics 
and a small displacement $\xi^a$ from it ($\xi^a \rightarrow L^a+\xi^a$). 
Up to second order in $\xi^a$ and $h_{\mu\nu}$, the effective action~\eqref{effective_action} is evaluated as
\begin{equation}
  S_\text{tot}=\frac{m}{2}\int dt \left( \delta_{ab} \frac{d\xi^{a}}{dt}\frac{d\xi^{b}}{dt}
    -
    2R^{(1)}_{0a0b} L^a \xi^b \right)  + \frac{1}{32\pi G_N} \int d^4x \, \Big[-\frac{1}{2}\partial_\alpha h_{\mu\nu} \partial^\alpha h^{\mu\nu}+\frac{1}{2} \partial_\alpha h \partial^\alpha h- \partial_\alpha h \partial_\beta h^{\alpha \beta}+\partial_\alpha h^{\alpha\mu} \partial_\beta h^{\beta}_{\mu}\Big].
\end{equation}
The term proportional to $m R^{(1)}_{0a0b} L^a L^b$ may be the source of gravitational field and the deviation $\xi^a$ can feel the sourced gravitational field. But the effect is a higher order contribution, and hence we ignored the term $m R^{(1)}_{0a0b} L^a L^b$ in the above effective action.
The energy--momentum tensor~\eqref{energy_mom_tensor1} can then be written as
\begin{equation}
    T^{\mu\nu}(x)
    =2m\int dt \,
    L^a \xi^b
    E_{0a0b}^{\mu\nu}
    \delta^4 (x-X(t)). \label{energy_momentum_tensor2}
\end{equation}
We assume the initial state of the gravitational field to be the classical vacuum,
$h_{\mu\nu}(t_i,\bm{x})=\dot h_{\mu\nu}(t_i,\bm{x})=0$, 
and define the Feynman--Vernon influence functional for $\xi^a$ as
\begin{equation}
    e^{i(S_0[\xi]-S_0[\underline{\xi}])+i S_{IF}[\xi,\underline{\xi}]} 
    = \int \mathcal{D} h_{\mu\nu} 
    \delta\!\left[\partial^{\mu}\!\left(h_{\mu\nu}-\frac{1}{2}\eta_{\mu\nu}h\right)\right] 
    e^{I_{\mathrm{CQ}}}
    \delta[h_{\mu\nu}(t_i)]
    \delta[\dot h_{\mu\nu}(t_i)],
\end{equation}
where $S_0[\xi]=m\int dt\,(\dot{\xi}^a)^2/2,$
and $\delta\!\left[\partial^{\mu}\!\left(h_{\mu\nu}-\frac{1}{2}\eta_{\mu\nu}h\right)\right]$
is the gauge-fixing condition.
After straightforward manipulations, we obtain
\begin{align}
    e^{i S_{IF}}
    &=
    \int \mathcal{D} h_{\mu\nu} \,
    \delta[\partial^{\mu}(h_{\mu\nu}-\tfrac{1}{2}\eta_{\mu\nu}h)] 
    \delta[h_{\mu\nu}(t_i)]
    \delta[\dot h_{\mu\nu}(t_i)]
    \notag\\
    &\quad \times
    e^{-im\int dt\, R^{(1)}_{0a0b} L^a(\xi^b-\underline{\xi}^b)}
    \notag\\
    &\quad \times
    e^{-\frac{1}{2}\!\int d^4x d^4y \,D_{\mu\nu\rho\sigma}(x,y)
    [T^{\mu\nu}(x)-\underline{T}^{\mu\nu}(x)]
    [T^{\rho\sigma}(y)-\underline{T}^{\rho\sigma}(y)]}
    \notag\\
    &\quad \times
    e^{-\frac{1}{2}\!\int d^4x d^4y \,
    N^{-1}_{\mu\nu\rho\sigma}(x,y)
    [G^{\mu\nu(1)}(x)-4\pi G_N (T^{\mu\nu}(x)+\underline{T}^{\mu\nu}(x))]
    [G^{\rho\sigma(1)}(y)-4\pi G_N (T^{\rho\sigma}(y)+\underline{T}^{\rho\sigma}(y))]}.
\end{align}

Here, we may insert the identity
\begin{equation}
    \int \mathcal{D}\chi_{\mu\nu}
    \,
    \delta\!\left[
    G^{(1)\mu\nu}
    -4\pi G_N (T^{\mu\nu}+\underline{T}^{\mu\nu})
    -\chi^{\mu\nu}
    \right]
    =1
\end{equation}
into the path integral. This leads to
\begin{align}
    e^{i S_{IF}}
    &=
    \int \mathcal{D}\chi_{\mu\nu}\,
    \exp\!\Bigg[
    - \frac{1}{2}\int d^4x d^4y \,
    D_{\mu\nu\rho\sigma}(x,y)
    \big(T^{\mu\nu}-\underline{T}^{\mu\nu}\big)_x
    \big(T^{\rho\sigma}-\underline{T}^{\rho\sigma}\big)_y
     -\frac{1}{2}\int d^4x d^4y \,
      N^{-1}_{\mu\nu\rho\sigma}(x,y)
      \chi^{\mu\nu}(x)\chi^{\rho\sigma}(y)\Bigg]
      \notag\\
    &\quad \times
    \int \mathcal{D} h_{\mu\nu}
    \,
    \delta\!\left[
    \partial^{\mu}\!\left(
    h_{\mu\nu}-\frac{1}{2}\eta_{\mu\nu}h
    \right)
    \right]
    \delta[h_{\mu\nu}(t_i,x)]
    \delta[\dot h_{\mu\nu}(t_i,x)]
    \delta\!\left[
    G^{(1)\mu\nu}
    -4\pi G_N (T^{\mu\nu}+\underline{T}^{\mu\nu})
    -\chi^{\mu\nu}
    \right]
    \notag\\
    &\quad \times
    \exp\!\left[
    -im\int dt\, R^{(1)}_{0a0b}
    L^a(\xi^b-\underline{\xi^b})
    \right].
\end{align}
The delta functionals in the path integral of $h_{\mu\nu}$ suggest that the path integral is evaluated by the solution of
the following Einstein--Langevin equations with the stochastic noise $\chi^{\mu\nu}$, 
\begin{equation}
    G^{(1)\mu\nu}
    =
    4\pi G_N \left(T^{\mu\nu}+\underline{T}^{\mu\nu}\right)
    + \chi^{\mu\nu},
    \label{Einstein_Langevan}
\end{equation}
under the initial conditions 
\begin{equation}
h_{\mu\nu}(t_i,\bm{x})=\dot h_{\mu\nu}(t_i,\bm{x})=0
 \label{initial_of_h}
\end{equation}
and the gauge-fixing conditions
\begin{equation}
\partial^{\mu}\!\Big(h_{\mu\nu}-\frac{1}{2}\eta_{\mu\nu}h\Big)=0. 
\label{gauge_condition}
\end{equation}

Let us solve the Einstein--Langevin equation by introducing
\begin{equation}
    \gamma^{\mu\nu}
    =
    h^{\mu\nu}
    -\frac{1}{2}\eta^{\mu\nu}h.
\end{equation}
The initial conditions are expressed by using this tensor as
\begin{equation}
    \gamma_{\mu\nu}(t_i,\bm{x})
    =
    \dot{\gamma}_{\mu\nu}(t_i, \bm{x})
    =
    0.
    \label{initial_gamma}
\end{equation}
and the gauge-fixing conditions are written as
\begin{equation}
\partial^{\mu} \gamma_{\mu\nu}=0. 
\label{gauge_condition_gamma}
\end{equation}
Rewriting the linearized Einstein tensor, 
\begin{align}
    G^{(1)\mu\nu}
    =
    \frac{1}{2}\Bigg[
    \frac{1}{2}\eta_{\mu\nu}
    \big(
    \partial_{\alpha}\partial_{\beta}h^{\alpha\beta}
    -\partial^2 h
    \big)
    - \frac{1}{2}\partial_{\beta}
    \big(
    \partial_{\mu}h^{\beta}_{\nu}
    +\partial_{\nu}h^{\beta}_{\mu}
    \big)
    + \frac{1}{2}\partial^2 h_{\mu\nu}
    + \frac{1}{2}\partial_{\mu}\partial_{\nu}h
    \Bigg],
\end{align}
by using $\gamma_{\mu\nu}$ and imposing the gauge conditions \eqref{gauge_condition_gamma}, we get
$G^{(1)\mu\nu}=\frac{1}{4}\partial^2 \gamma_{\mu\nu}$ and find 
the following Einstein--Langevin equation 
\begin{equation}
    \frac{1}{4}\partial^2 \gamma^{\mu\nu}
    - 4\pi G_N (T^{\mu\nu}+\underline{T}^{\mu\nu})
    - \chi^{\mu\nu}
    = 0.
\label{gamma_Einstein_Langevan}
\end{equation}
Noting that the initial conditions \eqref{initial_gamma} and using the retarded Green function $G_R$ together with Duhamel's principle, we can get the solution of  Eq.~\eqref{gamma_Einstein_Langevan} as
\begin{equation}
    \gamma_{\mu\nu}(x)
    =
    4\int_{t_i}^{\infty} dy^0
    \int_{\mathbb{R}^3} d^3y \,
    G_R(x-y)
    \left[
    4\pi G_N (T_{\mu\nu}(y)+\underline{T}_{\mu\nu}(y))
    +\chi_{\mu\nu}(y)
    \right].
\end{equation}
Using the trace relation 
$\gamma=\eta^{\mu\nu}\gamma_{\mu\nu}=\eta^{\mu\nu}(h_{\mu\nu}-\eta_{\mu\nu} h/2)=-h$, we can write the metric perturbation $h_{\mu\nu}$ in terms of $\gamma_{\mu\nu}$ as  $h^{\mu\nu}=\gamma^{\mu\nu}-\eta^{\mu\nu}\gamma/2$. 
Hence the solution of the Einstein-Langevin equation is give as  
\begin{align}
    h_{\mu\nu}(x)=h^T_{\mu\nu}(x)+h^\chi_{\mu\nu}(x),
\end{align}
where 
\begin{align}
    h^{T}_{\mu\nu}(x)
    &=
    4\int d^4y \,G_R(x-y) \,
    4\pi G_N
    \left[
    T_{\mu\nu}(y)
    +\underline{T}_{\mu\nu}(y)
    -\frac{1}{2}\eta_{\mu\nu}
    (T(y)+\underline{T}(y))
    \right],
    \\
    h^{\chi}_{\mu\nu}(x)
    &=
    4\int d^4y \,G_R(x-y)
    \left[
    \chi_{\mu\nu}(y)
    -\frac{1}{2}\eta_{\mu\nu}\chi(y)
    \right], 
\end{align}
with $T=\eta^{\mu\nu}T_{\mu\nu}$, $\underline{T}=\eta^{\mu\nu}\underline{T}_{\mu\nu}$ and $\chi=\eta^{\mu\nu}\chi_{\mu\nu}$.
The obtained solution shows that the gravitational field can be expressed as a sum of two contributions: the backreaction sourced by the quantum deviation, $h^T_{\mu\nu}$, and the gravitational field, $h^\chi_{\mu\nu}$ arising from the stochastic source $\chi_{\mu\nu}$.

Using this solution, the functional integral over $h_{\mu\nu}$ can be evaluated as
\begin{align}
    &\int \mathcal{D} h_{\mu\nu}
    \,
    \delta\!\left[
    \partial^{\mu}\!\left(
    h_{\mu\nu}-\frac{1}{2}\eta_{\mu\nu}h
    \right)
    \right]
    \delta[h_{\mu\nu}(t_i,x)]
    \delta[\dot h_{\mu\nu}(t_i,x)]
    \delta\!\left[
    G^{(1)\mu\nu}
    -4\pi G_N (T^{\mu\nu}+\underline{T}^{\mu\nu})
    -\chi^{\mu\nu}
    \right]
     e^{
    -im\int dt\, R^{(1)}_{0a0b}
    L^a(\xi^b-\underline{\xi^b})}
     \notag\\
    &= A
    \exp\!\left[
    -im\int dt\, R^{(1)*}_{0a0b}
    L^a(\xi^b-\underline{\xi^b})
    \right],
\end{align}
where $R^{(1)*}_{0a0b}$ is the spacetime curvature with $h_{\mu\nu}=h^T_{\mu\nu}+h^\chi_{\mu\nu}$, and $A$ is 
\begin{align}
    A=&\int \mathcal{D} h_{\mu\nu}
    \,
    \delta\!\left[
    \partial^{\mu}\!\left(
    h_{\mu\nu}-\frac{1}{2}\eta_{\mu\nu}h
    \right)
    \right]
    \delta[h_{\mu\nu}(t_i)]
    \delta[\dot h_{\mu\nu}(t_i)]
    \delta\!\left[
    G^{(1)\mu\nu}
    -4\pi G_N (T^{\mu\nu}+\underline{T}^{\mu\nu})
    -\chi^{\mu\nu}
    \right]. 
\end{align}
This $A$ is a constant and does not depend on $T^{\mu\nu}+\underline{T}^{\mu\nu}$ and $\chi^{\mu\nu}$ because the arguments of the delta functionals are linear in $h_{\mu\nu}$. Since the constant $A$ can be absorbed in the normalization, we can take $A=1$ in the following. Therefore, the influence functional can be written as
\begin{align}
    e^{i S_{IF}}
    &=
    \int \mathcal{D}\chi_{\mu\nu}\,
    \exp\!\Bigg[
    - \frac{1}{2}\int d^4x d^4y \,
    D_{\mu\nu\rho\sigma}(x,y)
    [T^{\mu\nu}(x)-\underline{T}^{\mu\nu}(x)]
    [T^{\rho\sigma}(y)-\underline{T}^{\rho\sigma}(y)]
    \Bigg]
      \notag\\
      &\quad \times
      \exp\!\Bigg[
      -\frac{1}{2}\int d^4x d^4y \,
      N^{-1}_{\mu\nu\rho\sigma}(x,y)
      \chi^{\mu\nu}(x)\chi^{\rho\sigma}(y)
      \Bigg]
    \exp\!\left[
    -im\int dt\, R^{(1)*}_{0a0b}
    L^a(\xi^b-\underline{\xi^b})
    \right].
    \label{SIF}
\end{align}
To perform the remaining functional integral over $\chi_{\mu\nu}$, it is necessary to make explicit the $\chi_{\mu\nu}$-dependence contained in $R^{(1)*}_{0a0b}$.
Since the curvature tensor is linear in the metric perturbation, which is formally given by $R^{(1)}_{\mu\alpha\nu\beta}=E_{\mu\alpha\nu\beta}^{\rho\sigma} h_{\rho\sigma}$, it is easy to identify the contribution from the stochastic source $\chi_{\mu\nu}$ as
\begin{align}
    R^{(1)*}_{\mu\alpha\nu\beta}
    =
    R^{(1)T}_{\mu\alpha\nu\beta}
    +
    R^{(1)\chi}_{\mu\alpha\nu\beta},
\end{align}
where 
$R^{(1)T}_{\mu\alpha\nu\beta}=E_{\mu\alpha\nu\beta}^{\rho\sigma} h^T_{\rho\sigma}$ and $R^{(1)\chi}_{\mu\alpha\nu\beta}=E_{\mu\alpha\nu\beta}^{\rho\sigma} h^\chi_{\rho\sigma}$. 
Since $R^{(1)\chi}_{\mu\alpha\nu\beta}$ is liner in $\chi_{\mu\nu}$, the functional integral over $\chi_{\mu\nu}$ given in \eqref{SIF} is a Gaussian functional integral. Hence, the influence functional is calculated as
\begin{align}
    e^{iS_{IF}}
    &=
    \exp\!\Bigg[-im\int dt\, R^{(1)T}_{0a0b} L^a(\xi^b-\underline{\xi^b}) - \frac{1}{2}\int d^4x d^4y \,D_{\mu\nu\rho\sigma}(x,y)
    [T^{\mu\nu}(x)-\underline{T}^{\mu\nu}(x)]
    [T^{\rho\sigma}(y)-\underline{T}^{\rho\sigma}(y)]\bigg]
      \notag\\
      &\qquad \times
      \exp\!\Bigg[-\frac{1}{2}m^2\!\int dt dt' \,
    \braket{R^{(1)\chi}_{0a0b}(t) \,R^{(1)\chi}_{0c0d}(t')}L^aL^c (\xi^b(t)-\underline{\xi^b}(t)) (\xi^d(t')-\underline{\xi^d}(t')) \bigg].
\end{align}

We explicitly rewrite the influence functional by the deviation vector and derive the Langevin equation for the deviation. Substituting the stress tensor \eqref{energy_momentum_tensor2}, we first obtain
\begin{align}
    &\int d^4x d^4y \,
    D_{\mu\nu\rho\sigma}(x,y)
    \big(T^{\mu\nu}(x)-\underline{T}^{\mu\nu}(x)\big)
    \big(T^{\rho\sigma}(y)-\underline{T}^{\rho\sigma}(y)\big)
    =
    \int_{t_i}^{t_f} dt\,dt' \,
    \Delta^D_{abcd}(t,t')
    L^a L^c
    (\xi^b(t)-\underline{\xi^b}(t))
    (\xi^d(t')-\underline{\xi^d}(t')),
\end{align}
where 
\begin{equation}
    \Delta^D_{abcd}(t,t')
    =
    4m^2
    E_{0a0b}^{x,\mu\nu}
    E_{0c0d}^{y,\rho\sigma}
    D_{\mu\nu\rho\sigma}(x,y)
    \big|_{x^\mu=X^\mu(t),\,y^\mu=X^\mu(t')}.
\end{equation}
Here the superscripts $x$ and $y$ indicate that the derivatives act on $x$ and $y$, respectively.
Next, for the curvature $R^{(1)T}_{0a0b}$, we get
\begin{align}
    R^{(1)T}_{0a0b}(t)
    &=
    -\frac{1}{2m}
    \int^t_{t_i} dt' \,
    \Sigma_{abcd}(t,t')
    L^c
    (\xi^d(t')+\underline{\xi^d}(t')),
\end{align}
with the dissipation kernel
\begin{align}
    \Sigma_{abcd}(t,t')
    =
    -64\pi G_N m^2
    E^{x,\mu\nu}_{0a0b}
    \left(
    E^{y}_{0c0d\,\mu\nu}
    - \frac{1}{2}\eta^{\alpha\beta}
    E^{y}_{0c0d\,\alpha\beta}
    \eta_{\mu\nu}
    \right)
    G_R(x-y)
    \big|_{x^\mu=X^\mu(t),\,y^\mu=X^\mu (t')}.
\label{dissipation_kernel}
\end{align}
Furthermore, the stochastic curvature correlation becomes
\begin{align}
\Delta^N_{abcd}(t,t')
&=m^2
    \big\langle
    R^{(1)\chi}_{0a0b}(t)
    R^{(1)\chi}_{0c0d}(t')
    \big\rangle
    \notag\\
    &=
    16m^2
    \int d^4z d^4w
    \,
    E^{x,\mu\nu}_{0a0b}
    G_R(x-z)
    E^{y,\rho\sigma}_{0c0d}
    G_R(y-w)\big|_{x^\mu=X^\mu(t),y^\mu=X^\mu(t')}
    \notag\\
    &\quad \times
    (\delta^{\alpha}_{\mu}\delta^{\beta}_{\nu}
    -\tfrac{1}{2}\eta_{\mu\nu}\eta^{\alpha\beta})
    (\delta^{\lambda}_{\rho}\delta^{\kappa}_{\sigma}
    -\tfrac{1}{2}\eta_{\rho\sigma}\eta^{\lambda\kappa})
    N_{\alpha\beta\lambda\kappa}(z,w). 
\end{align}
Using the above kernels $\Delta^D_{abcd}$, $\Sigma_{abcd}$ and $\Delta^N_{abcd}$, we get the following expression of the influence functional, 
\begin{align}
    e^{iS_{IF}[\xi,\underline{\xi}]}
    &=
    \exp\Bigg[
    \frac{i}{2}
    \int^{t_f}_{t_i} dt
    \int^t_{t_i} dt' \,
    \Sigma_{abcd}(t,t')
    L^a L^c
    (\xi^b(t)-\underline{\xi^b}(t))
    (\xi^d(t')+\underline{\xi^d}(t'))
    \Bigg]
    \notag\\
    &\quad \times
    \exp\Bigg[
    -\frac{1}{2}
    \int^{t_f}_{t_i} dt\,dt'
    \left(
    \Delta^D_{abcd}(t,t')
    + \Delta^N_{abcd}(t,t')
    \right)
    L^a L^c
    (\xi^b(t)-\underline{\xi^b}(t))
    (\xi^d(t')-\underline{\xi^d}(t'))
    \Bigg].
\end{align}
According to the standard procedure to read out the Langevin equation from the action
\(
S_0[\xi] - S_0[\underline{\xi}] + S_{IF}[\xi,\underline{\xi}]
\) \cite{Calzetta2008}, 
we can derive the Langevin equation
\begin{align}
    m\frac{d^2 \xi^a}{dt^2}
    - \zeta^a(t)
    = 0,
\end{align}
where we neglected the dissipation term associated with the kernel $\Sigma_{abcd}$. 
This dissipation comes from  gravitational wave radiation from the motion of the deviation, and hence it can be negligible by assuming a sufficiently small mass $m$. 

\section{Derivation of the power spectral density $(S^h_{x,\text{org}})^2$}
\label{Ap:Derivation_spectrum_ori}

In this section, we show how the power spectral density $(S^h_{x,\text{org}})^2$,  Eq.\eqref{Spectrum_Oppenheim_original}, is obtained from the decoherence kernel $D_{\mu\nu\rho\sigma}$ and the noise kernel $N_{\mu\nu\rho\sigma}$, which are given in \eqref{Op_Deco} and \eqref{Op_Noise}, respectively.
To this end, we first evaluate the correlation function $\Delta^D_{abcd}$ in \eqref{DeltaD} and the contribution of the decoherence kernel, $(S^D_x)^2$, defined in \eqref{SDN}. 
Substituting Eq. \eqref{Op_Deco} into Eq. \eqref{DeltaD}, we have
\begin{align}
    &\Delta^D_{abcd}(t,t') 
    \notag 
    \\
    &\quad =
    4m^2 E_{0a0b}^{x,\mu\nu} E_{0c0d}^{y,\rho\sigma} D_{\mu\nu\rho\sigma}(x,y)|_{x^\mu=X^\mu(t), y^\mu=X^\mu(t')} \notag\\
    &\quad =\frac{m^2 D_0^{\mathrm{\,org}}}{2} \int \frac{d^4k}{(2\pi)^4} \bigg[(2-2\beta)k_ak_bk_ck_d  -(k^0)^2\{ k_ak_c\delta_{bd} +k_ck_b\delta_{ad} +k_ak_d\delta_{bc} +k_bk_d\delta_{ac}- 2\beta (k_ck_d\delta_{ab} + k_ak_b \delta_{cd})\} \notag 
    \\
    &\quad \quad+(k^0)^4(\delta_{ac}\delta_{bd}+\delta_{ad}\delta_{bc}-2\beta \delta_{ab}\delta_{cd}) \bigg] e^{-ik^0(t-t')}.
    \label{c1}
\end{align}
In this step, we used the fact that the delta function can be written as $\delta^4(x-y)=\int \frac{d^4k}{(2\pi)^4}e^{ik_{\mu}(x^{\mu}-y^{\mu})}$, and $x^\mu=X^\mu(t)=(t,0,0,0),\, y^\mu=X^\mu(t')=(t',0,0,0)$.
Here, for the integration over the spatial components of $k_a$, we introduce a UV cutoff as the inverse of the mean separation length $L$. The integral formulas
\begin{align}
    \int_{|k|\le 1/L} d^3k\, k_ak_b=\frac{4\pi}{15L^5}\delta_{ab}, 
    \quad \int_{|k|\le 1/L} d^3k\,k_ak_bk_ck_d=\frac{4\pi}{105L^7}(\delta_{ab}\delta_{cd}+\delta_{ac}\delta_{bd}+\delta_{ad}\delta_{bc}),\label{intC}
\end{align}
reduce Eq.~\eqref{c1} to
\begin{align}
    \Delta^D_{abcd}(t,t')
    &=
     \frac{m^2 D_0^{\mathrm{\,org}}}{2(2\pi)^4} \int dk^0 \bigg[ \frac{4\pi}{3L^3}(k^0)^4\{\delta_{ac}\delta_{bd}+\delta_{ad}\delta_{bc}-2\beta \delta_{ab}\delta_{cd}\}+\frac{8\pi}{15L^5}(k^0)^2 (2\beta\delta_{ab}\delta_{cd}-\delta_{ac}\delta_{bd}-\delta_{ad}\delta_{bc})  \notag\\
    &\quad +\frac{8\pi}{105L^7}(1-\beta)(\delta_{ab}\delta_{cd}+\delta_{ac}\delta_{bd}+\delta_{ad}\delta_{bc})\bigg] e^{-ik^0(t-t')}.
\end{align}
According to \eqref{SDN}, this $\Delta^D_{abcd}$ yields 
\begin{align}
(S^D_{x,\text{org}})^2
=\frac{1}{m^2\omega^4}\int dt \,e^{i\omega t} \,\Delta^D_{xxxx} (t,0)
=\frac{D_0^{\mathrm{\,org}}}{\pi^2}(1-\beta)\bigg[ \frac{1}{3L^3}-\frac{2}{15L^5\omega^2}+ \frac{1}{35L^7\omega^4}\bigg].
    \label{c2}
\end{align}

Next, we calculate the correlation function $\Delta^N_{abcd}$ in Eq.\eqref{DeltaN} and the contribution of the noise kernel, $(S^N_x)^2$, given in \eqref{SDN}.  
Using the retarded Green function  Eq.~\eqref{R_Green_function} and substituting Eq.\eqref{Op_Noise} into Eq. \eqref{DeltaN}, we obtain
\begin{align}
    &\Delta^N_{abcd}(t,t') 
    \notag \\
    &\quad =\frac{1024\pi^2m^2G_N^2}{D_0^{\mathrm{\,org}}}\int \frac{d^4k}{(2\pi)^4} \frac{e^{-ik^0(t-t')}}{|-(k^0+i\epsilon)^2+\boldsymbol{k}^2|^2} \bigg[\frac{1}{2}(k^0)^4 \Big(\delta_{ac}\delta_{bd} +\delta_{ad}\delta_{bc}+\frac{2\beta}{1-4\beta}\delta_{ab}\delta_{cd}\Big)\notag\\
    &\quad
     -\frac{1}{2}(k^0)^2\Big(k_ak_c\delta_{bd} +k_bk_c\delta_{ad}+k_bk_d\delta_{ac} + k_ak_d\delta_{bc} +\frac{2\beta}{1-4\beta} (k_ak_b\delta_{cd} + k_ck_d\delta_{ab})\Big)
    +\frac{1-3\beta}{1-4\beta}k_ak_bk_ck_d  \bigg].
\end{align}
Here, by applying the integral formulas,
\begin{align}
    &\int_{|k|\le 1/L}d^3k\, \frac{1}{|-(k^0+i\epsilon)^2+\boldsymbol{k}^2|^2} \notag\\
    &\qquad=
    4\pi\frac{(k^0+i\epsilon)\mathrm{Arccot}[L(\epsilon-ik^0)] + (k^0-i\epsilon)\mathrm{Arccot}[L(\epsilon+ik^0)]}{4k^0\epsilon}, 
    \label{intC2}\\
    &\int_{|k|\le 1/L}d^3k\, \frac{k_ak_b }{|-(k^0+i\epsilon)^2+\boldsymbol{k}^2|^2} \notag\\
    &\qquad=
    \frac{4\pi}{3} \frac{4k^0\epsilon-iL(\epsilon-ik^0)^3\mathrm{Arccot}[L(\epsilon-ik^0)] + iL(\epsilon+ik^0)^3\mathrm{Arccot}[L(\epsilon+ik^0)]}{4k^0L\epsilon}\delta_{ab}, 
    \label{intC3}\\
    &\int_{|k|\le 1/L}d^3k\, \frac{ k_ak_bk_ck_d}{|-(k^0+i\epsilon)^2+\boldsymbol{k}^2|^2} \notag\\
    &\qquad=
    \frac{4\pi}{15} \bigg[ \frac{1}{3L^3} + \frac{2(k^0-\epsilon)(k^0+\epsilon)}{L} + \frac{(k^0+i\epsilon)^5\mathrm{Arccot}[L(\epsilon-ik^0)] + (k^0-i\epsilon)^5\mathrm{Arccot}[L(\epsilon+ik^0)]}{4k^0\epsilon} \bigg] \notag\\& \quad \qquad \times(\delta_{ab}\delta_{cd}+\delta_{ac}\delta_{bd}+\delta_{ad}\delta_{bc}),
    \label{intC4}
\end{align}
the correlation function $\Delta^N_{abcd}$ is 
\begin{align}
    &\Delta^N_{abcd}(t,t') 
    \notag\\
      &=\frac{64m^2 G_N^2}{\pi^2 D_0^{\mathrm{\,org}}} \int dk^0 \bigg[ 2\pi (k^0)^4\left(\delta_{ac}\delta_{bd} + \delta_{ad}\delta_{bc}+\frac{2\beta}{1-4\beta}\delta_{ab}\delta_{cd}\right)\frac{(k^0+i\epsilon)\mathrm{Arccot}[L(\epsilon-ik^0)] + (k^0-i\epsilon)\mathrm{Arccot}[L(\epsilon+ik^0)]}{4k^0\epsilon}  \notag\\&\quad - \frac{4\pi}{3} (k^0)^2\left(\delta_{ac}\delta_{bd} + \delta_{ad}\delta_{bc}+\frac{2\beta}{1-4\beta}\delta_{ab}\delta_{cd}\right) \frac{4k^0\epsilon-iL(\epsilon-ik^0)^3\mathrm{Arccot}[L(\epsilon-ik^0)] + iL(\epsilon+ik^0)^3\mathrm{Arccot}[L(\epsilon+ik^0)]}{4k^0 L\epsilon}  \notag\\ &\quad +\frac{4\pi}{15}\frac{1-3\beta}{1-4\beta} \bigg( \frac{1}{3L^3} + \frac{2(k^0-\epsilon)(k^0+\epsilon)}{L} + \frac{(k^0+i\epsilon)^5\mathrm{Arccot}[L(\epsilon-ik^0)] + (k^0-i\epsilon)^5\mathrm{Arccot}[L(\epsilon+ik^0)]}{4k^0\epsilon} \bigg) \notag\\&\qquad  \times(\delta_{ab}\delta_{cd}+\delta_{ac}\delta_{bd}+\delta_{ad}\delta_{bc}) \bigg] e^{-ik^0 (t-t')}.
\end{align}
Following Eqs.\eqref{SDN}, the contribution from the noise kernel is yielded as
\begin{align}
    (S^N_{x,\text{org}})^2
    &=
    \frac{1}{m^2\omega^4} \int dt\,e^{i\omega t} \Delta^N_{xxxx}(t,0) \notag\\
    \quad &=
    \frac{512G_N^2}{ D_0^{\mathrm{\,org}}}  \frac{1-3\beta}{1-4\beta}\bigg[  \frac{(\omega+i\epsilon)\mathrm{Arccot}[L(\epsilon-i\omega)] + (\omega-i\epsilon)\mathrm{Arccot}[L(\epsilon+i\omega)]}{4\omega\epsilon}  \notag\\
    &\quad - \frac{4\omega\epsilon-iL(\epsilon-i\omega)^3\mathrm{Arccot}[L(\epsilon-i\omega)] + iL(\epsilon+i\omega)^3\mathrm{Arccot}[L(\epsilon+i\omega)]}{6\omega^3 L\epsilon}  \notag\\
    &\quad + \bigg( \frac{1}{15L^3\omega^4} + \frac{2(\omega^2-\epsilon^2)}{5L\omega^4} + \frac{(\omega+i\epsilon)^5\mathrm{Arccot}[L(\epsilon-i\omega)] + (\omega-i\epsilon)^5\mathrm{Arccot}[L(\epsilon+i\omega)]}{20\omega^5\epsilon} \bigg)\Big].
    \label{c3}
\end{align}

According to Eq.\eqref{SD+SN}, the power spectral density is given by adding Eqs.~\eqref{c2} and \eqref{c3} as 
\begin{align}
    (S^h_{x,\text{org}})^2
    &\notag=
    (S^D_{x,\text{org}})^2 + (S^N_{x,\text{org}})^2 \notag \\
    &=
    \frac{D_0^{\mathrm{\,org}}}{\pi^2}(1-\beta) \left( \frac{1}{3L^3} - \frac{2}{15L^5\omega^2} + \frac{1}{35L^7\omega^4} \right) \\ \notag
    & \quad + \frac{512}{D_0^{\mathrm{\,org}} m_p^4} \frac{1-3\beta}{1-4\beta} \left[ \frac{(\omega+i\epsilon)\mathrm{Arccot}[L(\epsilon-i\omega)] + (\omega-i\epsilon)\mathrm{Arccot}[L(\epsilon+i\omega)]}{4\epsilon\omega}\right. \\\notag
    &\qquad -   \frac{4\omega\epsilon-iL(\epsilon-i\omega)^3\mathrm{Arccot}[L(\epsilon-i\omega)] + iL(\epsilon+i\omega)^3\mathrm{Arccot}[L(\epsilon+i\omega)]}{6 L\epsilon\omega^3} \\
    &\qquad \left. +  \left( \frac{1}{15L^3 \omega^4} + \frac{2(\omega^2 - \epsilon^2)}{5L\omega^4} + \frac{(\omega+i\epsilon)^5\mathrm{Arccot}[L(\epsilon-i\omega)] + (\omega-i\epsilon)^5\mathrm{Arccot}[L(\epsilon+i\omega)]}{20\epsilon\omega^5} \right) \right],
\end{align}
where $m^2_p=1/G_N$.

\section{Derivation of the power spectral density $(S^h_{x,\text{Bia}})^2$}
\label{Ap:Derivation_spectrum_Einstein}

In this section, we derive the power spectral density $(S^h_{x,\text{Bia}})^2$ given in \eqref{Spectrum_white_Ein} following the same procedure as in Appendix \ref{Ap:Derivation_spectrum_ori}.
We first evaluate the terms involving the decoherence kernel. The decoherence kernel $D_{\mu\nu\rho\sigma}$ defined by Eq.\eqref{Ein_Deco} gives the correlation function $\Delta^D_{abcd}$ as
\begin{align}
    \Delta^D_{abcd}(t,t')
    &=
    4m^2 D_0^{\mathrm{\,Bia}} \Big\{E_{0a0b}^{x,\mu\nu} E_{0c0d}^{y,\rho\sigma} \int \frac{d^4p}{(2\pi)^4} e^{ip^\mu(x_\mu-y_\mu)} \Big[ (\eta_{\mu\rho} - \frac{p_{\mu}p_{\rho}}{p^2}) (\eta_{\sigma\nu} - \frac{p_{\sigma}p_{\nu}}{p^2}) + (\eta_{\mu\sigma} - \frac{p_{\mu}p_{\sigma}}{p^2}) (\eta_{\rho\nu} - \frac{p_{\rho}p_{\nu}}{p^2})\notag\\
    &\qquad
    -\frac{2}{3} (\eta_{\mu\nu} - \frac{p_{\mu}p_{\nu}}{p^2}) (\eta_{\rho\sigma} - \frac{p_{\rho}p_{\sigma}}{p^2})\Big] \Big \}_{x^\mu=X^\mu(t), y^\mu=X^\mu(t')} \notag\\
    &=
    4m^2 D_0^{\mathrm{\,Bia}} \int \frac{d^4p}{(2\pi)^4} e^{-ip^0(t-t')} \bigg[ (p^0)^4\bigg(\delta_{ac}\delta_{bd} + \delta_{ad}\delta_{bc} -\frac{2}{3}\delta_{ab}\delta_{cd}\bigg) +\frac{4}{3}p_ap_bp_cp_d \notag\\
    &\qquad-
    (p^0)^2 \bigg(p_ap_c\delta_{bd} + p_ap_d\delta_{bc} + p_bp_c \delta_{ad} + p_bp_d\delta_{ac} -\frac{2}{3}(p_ap_b\delta_{cd}+p_cp_d\delta_{ab})\bigg) \bigg].
    \label{d0}
\end{align}
Here, by using the integral formulas in Eqs.~\eqref{intC}, the contribution $(S^D_{x,\text{Bia}})^2$ defined in Eqs.\eqref{SDN} is simplified as 
\begin{align}
    (S^D_{x,\text{Bia}})^2
    =
    \frac{1}{m^2\omega^4} \int dt\,e^{i\omega t} \Delta^D_{xxxx}(t,0) =\frac{8D_0^{\mathrm{\,Bia}}}{\pi^2}\left( \frac{1}{9L^3} -\frac{2}{45L^5 \omega^2} +\frac{1}{105L^7\omega^4} \right) ,
    \label{d1}
\end{align}
where the inverse of the mean separation $L$ was adopted as the UV cutoff parameter again. 

Next, we calculate the terms involving the noise kernel.   
Using the same retarded Green function \eqref{R_Green_function}, we can get the following form of the correlation function $\Delta^N_{abcd}$:
\begin{align}
    \Delta^N_{abcd}(t,t')
    &=
    \frac{16m^2 (4\pi G_N)^2}{D_0^{\mathrm{\,Bia}}}\int \frac{d^4p}{(2\pi)^4}\, \frac{ e^{-ip^0(t-t')}}{|-(p^0+i\epsilon)^2+\boldsymbol{p}^2|^2}   \bigg[ (p^0)^4\bigg(\delta_{ac}\delta_{bd} + \delta_{ad}\delta_{bc} -\frac{2}{3}\delta_{ab}\delta_{cd}\bigg) +\frac{4}{3}p_ap_bp_cp_d \notag\\
    &\quad-
    (p^0)^2 \bigg(p_ap_c\delta_{bd} + p_ap_d\delta_{bc} + p_bp_c \delta_{ad} + p_bp_d\delta_{ac} -\frac{2}{3}(p_ap_b\delta_{cd}+p_cp_d\delta_{ab})\bigg)  \bigg].
    \label{eq:Delta_Env}
\end{align}
Here, regularizing the UV divergence in the three-momentum integral by $1/L$ and applying the integrals given in Eqs.\eqref{intC2},\eqref{intC3} and \eqref{intC4},
we get the noise contribution $(S^N_{x,\text{Bia}})^2$ to the power spectral density as
\begin{align}
    (S^N_{x,\text{Bia}})^2
    &=
    \frac{1}{m^2\omega^4} \int dt\,e^{i\omega t} \Delta^N_{xxxx}(t,0) \notag\\
    \quad &=
    + \frac{128}{D_0^{\mathrm{\,Bia}} m_p^4} \left[  \frac{(\omega +i\epsilon)\mathrm{Arccot}[L(\epsilon-i\omega)] + (\omega -i\epsilon)\mathrm{Arccot}[L(\epsilon+i\omega)]}{3\epsilon \omega}\right. \\\notag
    &\qquad - 2 \frac{4\epsilon \omega - iL (\epsilon - i\omega)^3 \mathrm{Arccot}[L(\epsilon-i\omega)] + iL (\epsilon + i\omega)^3 \mathrm{Arccot}[L(\epsilon+i\omega)]}{9L\epsilon \omega^3} \\
    &\qquad \left. + \frac{4}{15}  \left( \frac{1}{3L^3 \omega^4} + \frac{2(\omega^2 - \epsilon^2)}{L\omega^4} + \frac{(\omega + i\epsilon)^5 \mathrm{Arccot}[L(\epsilon-i\omega)] +  (\omega - i\epsilon)^5 \mathrm{Arccot}[L(\epsilon+i\omega)]}{4\epsilon \omega^5} \right) \right].
    \label{d2}
\end{align}
The sum of $(S^D_{x,\text{Bia}})^2$ and $(S^N_{x,\text{Bia}})^2$ gives the following power spectral density,
\begin{align}
    (S^h_{x,\text{Bia}})^2&\notag
    =
    (S^D_{x,\text{Bia}})^2 + (S^N_{x,\text{Bia}})^2 \notag\\
    &= \frac{8D_0^{\mathrm{\,Bia}}}{\pi^2}\left( \frac{1}{9L^3} -\frac{2}{45L^5 \omega^2} +\frac{1}{105L^7\omega^4} \right) \\ \notag
    & \quad + \frac{128}{D_0^{\mathrm{\,Bia}} m_p^4} \left[  \frac{(\omega +i\epsilon)\mathrm{Arccot}[L(\epsilon-i\omega)] + (\omega -i\epsilon)\mathrm{Arccot}[L(\epsilon+i\omega)]}{3\epsilon \omega}\right. \\\notag
    &\qquad - 2 \frac{4\epsilon \omega - iL (\epsilon - i\omega)^3 \mathrm{Arccot}[L(\epsilon-i\omega)] + iL (\epsilon + i\omega)^3 \mathrm{Arccot}[L(\epsilon+i\omega)]}{9L\epsilon \omega^3} \\
    &\qquad \left. + \frac{4}{15}  \left( \frac{1}{3L^3 \omega^4} + \frac{2(\omega^2 - \epsilon^2)}{L\omega^4} + \frac{(\omega + i\epsilon)^5 \mathrm{Arccot}[L(\epsilon-i\omega)] +  (\omega - i\epsilon)^5 \mathrm{Arccot}[L(\epsilon+i\omega)]}{4\epsilon \omega^5} \right) \right],
\end{align}
and this is nothing but \eqref{Spectrum_white_Ein}.

\section{Derivation of the power spectral density $(S^h_{x,\text{env}})^2$ and its minimum}
\label{Ap:Derivation_spectrum_environment}

In this section, we derive the power spectral density following the same procedure as in Appendices \ref{Ap:Derivation_spectrum_ori} and \ref{Ap:Derivation_spectrum_Einstein}. We begin with the derivation of Eq.~\eqref{Spectrum_Environmental_Oppenheim}.
In this model, the decoherence kernel is given by Eq.~\eqref{Ein_Deco} multiplied by $\theta(-p^2-4\mu^2)=\theta(p_0^2-\boldsymbol{p}^2-4\mu^2)$, so we can simply insert the step function into \eqref{d0} to get the following correlation, 
\begin{align}
    \Delta^D_{abcd}(t,t')
    &=
    D_0^{\mathrm{\,env}} \int \frac{d^4p}{(2\pi)^4} e^{-ip^0(t-t')}\, \theta((p^0)^2-\boldsymbol{p}^2-4\mu^2) \bigg[ (p^0)^4\bigg(\delta_{ac}\delta_{bd} + \delta_{ad}\delta_{bc} -\frac{2}{3}\delta_{ab}\delta_{cd}\bigg) +\frac{4}{3}p_ap_bp_cp_d \notag\\
    &\qquad-
    (p^0)^2 \bigg(p_ap_c\delta_{bd} + p_ap_d\delta_{bc} + p_bp_c \delta_{ad} + p_bp_d\delta_{ac} -\frac{2}{3}(p_ap_b\delta_{cd}+p_cp_d\delta_{ab})\bigg) \bigg].
    \label{e1}
\end{align}
Here, from the step function $\theta((p^0)^2-\bm{p}^2-4\mu^2)$, the integration range should be $(p^0)^2-\bm{p}^2-4\mu^2>0$. While the integration range of $p^0$ is assumed to be from $2\mu$ to $\infty$ because we focus on the positive frequency $\omega$ to get the power spectral density. 
We then get the following integral formulas 
\begin{align}
    &\int d^3p \, \theta((p^0)^2-\bm{p}^2-4\mu^2) \, =\frac{4\pi}{3} [(p^0)^2-4\mu^2]^{\frac{3}{2}} \theta(p^0-2\mu), \label{intE1}
    \\
    &\int d^3p \, \theta((p^0)^2-\bm{p}^2-4\mu^2) \, p_ap_b=\frac{4\pi}{15}[(p^0)^2-4\mu^2]^{\frac{5}{2}} \theta(p^0-2\mu)\, \delta_{ab} , \label{intE2}\\
   &\int d^3p\,\theta((p^0)^2-\bm{p}^2-4\mu^2) p_a p_b p_c p_d =\frac{4\pi}{105}[(p^0)^2-4\mu^2]^{\frac{7}{2}} \theta(p^0-2\mu)(\delta_{ab}\delta_{cd}+\delta_{ac}\delta_{bd}+\delta_{ad}\delta_{bc}).\label{intE3}
\end{align}
Using these relations to proceed with the calculation, we obtain the decoherence contribution as 
\begin{align}
    (S^D_{x,\text{env}})^2
    &=
    \frac{1}{m^2\omega^4} \int dt\,e^{i\omega t} \Delta^D_{xxxx}(t,0)\notag\\
    \quad &=
    \frac{64D_0^{\mathrm{\,env}}}{315\pi^2} \theta(\omega -2\mu) [\omega^2-4\mu^2]^{\frac{3}{2}}
    \frac{3\omega^4+4\mu^2 \omega^2+6\mu^4}{\omega^4}.
    \label{e5}
\end{align}

Next, we calculate the contribution from the noise kernel. The procedure is similar to the above one. Inserting the step function $\theta ((p^0)^2-\bm{p}^2-4\mu^2)$ into \eqref{eq:Delta_Env}, we have the corresponding $\Delta^N_{abcd}$ in this model as 
\begin{align}
&\Delta^N_{abcd}(t,t') \notag\\
&\quad =
 \frac{16m^2 (4\pi G_N)^2}{D_0^{\mathrm{\,Bia}}}\int \frac{d^4p}{(2\pi)^4}\, \frac{ e^{-ip^0(t-t')} \theta((p^0)^2-\bm{p}^2-4\mu^2)}{|-(p^0)^2+\boldsymbol{p}^2|^2}   \bigg[ (p^0)^4\bigg(\delta_{ac}\delta_{bd} + \delta_{ad}\delta_{bc} -\frac{2}{3}\delta_{ab}\delta_{cd}\bigg) +\frac{4}{3}p_ap_bp_cp_d \notag\\
    &\quad-
    (p^0)^2 \bigg(p_ap_c\delta_{bd} + p_ap_d\delta_{bc} + p_bp_c \delta_{ad} + p_bp_d\delta_{ac} -\frac{2}{3}(p_ap_b\delta_{cd}+p_cp_d\delta_{ab})\bigg)  \bigg],
\end{align}
where we took the limit $\epsilon \rightarrow 0$
since the four-momentum integral is performed for $-p^2>4\mu^2$ and the integrands are not singular at $p^2=0$. 
Applying the integral formulas, 
\begin{align}
    &\int d^3p \, \frac{\theta((p^0)^2-\bm{p}^2-4\mu^2)}{|-(p^0)^2+\bm{p}^2|^2} \, =\frac{\pi}{2} \Big\{\frac{\sqrt{(p^0)^2-4\mu^2}}{\mu^2}-\frac{4}{p^0} \text{Arccoth}\Big[\frac{p^0}{\sqrt{(p^0)^2-4\mu^2}}\Big] \Big\}\theta(p^0-2\mu), \label{intE7}
    \\
    &\int d^3p \, \frac{\theta((p^0)^2-\bm{p}^2-4\mu^2)}{|-(p^0)^2+\bm{p}^2|^2} \, p_ap_b=\frac{4\pi}{3} \Big\{\frac{(p^0)^2+8\mu^2}{8\mu^2}\sqrt{(p^0)^2-4\mu^2}-\frac{3p^0}{2} \text{Arccoth}\Big[\frac{p^0}{\sqrt{(p^0)^2-4\mu^2}}\Big] \Big\}\theta(p^0-2\mu)\, \delta_{ab} , \label{intE8}\\
   &\int d^3p\, \frac{\theta((p^0)^2-\bm{p}^2-4\mu^2)}{|-(p^0)^2+\bm{p}^2|^2} p_a p_b p_c p_d 
   \notag
   \\
   &\quad =\frac{4\pi}{15}\Big\{\frac{3(p^0)^4+56(p^0)^2 \mu^2-32\mu^4}{24\mu^2}\sqrt{(p^0)^2-4\mu^2}-\frac{5(p^0)^3}{2} \text{Arccoth}\Big[\frac{p^0}{\sqrt{(p^0)^2-4\mu^2}}\Big] \Big\} \theta(p^0-2\mu) \notag
   \\
   &\quad \quad \times  (\delta_{ab}\delta_{cd}+\delta_{ac}\delta_{bd}+\delta_{ad}\delta_{bc}), \label{intE9}
\end{align}
we can evaluate the noise contribution as  
\begin{align}
    (S^N_{x,\text{env}})^2
    &=
    \frac{1}{m^2\omega^4} \int dt\,e^{i\omega t} \Delta^N_{xxxx}(t,0)\notag\\
    \quad &=
    \frac{512}{45m^4_p  D_0^{\mathrm{\,env}}} \theta(\omega -2\mu)[\omega^2-4\mu^2]^{\frac{3}{2}}\frac{\omega^2+\mu^2}{\mu^2 \omega^4} .
    \label{e10}
\end{align}
As in Appendices \ref{Ap:Derivation_spectrum_ori} and \ref{Ap:Derivation_spectrum_Einstein}, by adding Eqs.~\eqref{e5} and \eqref{e10} and simplifying, the power spectral density is yielded as
\begin{align}
    (S^h_{x,\text{env}})^2
    &\notag=
    (S^D_{x,\text{env}})^2 + (S^N_{x,\text{env}})^2 \notag\\
    &=\frac{64D_0^{\mathrm{\,env}}}{315\pi^2} \theta(\omega -2\mu) [\omega^2-4\mu^2]^{\frac{3}{2}}
    \frac{3\omega^4+4\mu^2 \omega^2+6\mu^4}{\omega^4}+ \frac{512}{45m^4_p  D_0^{\mathrm{\,env}}} \theta(\omega -2\mu)[\omega^2-4\mu^2]^{\frac{3}{2}}\frac{\omega^2+\mu^2}{\mu^2 \omega^4}.
\end{align}

Finally, we derive Eq.~\eqref{mini_ENmodel} and calculate the minimum of power spectral density for any spectra of the noise and decoherence kernels, that is, $N(p)$ and $D(p)$. The lower bound Eq.\eqref{mini_ENtwo} for the sum of the correlations is evaluated as 
\begin{align}
&\Delta^D_{abcd} + \Delta^N_{abcd} \notag \\
&\quad=4m^2 E_{0a0b}^{x,\mu\nu} E_{0c0d}^{y,\rho\sigma} D_{\mu\nu\rho\sigma}(x,y)|_{x^\mu=X^\mu(t), y^\mu=X^\mu(t')} \notag \\
&\quad+16m^2 \int^t_{-\infty} d^4 z  \int^{t'}_{-\infty} d^4 w  E_{0a0b}^{x,\mu\nu} G_R(x-z) E_{0c0d}^{y,\rho\sigma} G_R(y-w)|_{x^\mu=X^\mu(t), y^\mu=X^\mu(t')} \notag\\
&\quad\times (\delta^{\alpha}_{\mu} \delta^{\beta}_{\nu} - \frac{1}{2}\eta_{\mu\nu}\eta^{\alpha\beta}) (\delta^{\lambda}_{\rho} \delta^{\kappa}_{\sigma} - \frac{1}{2}\eta_{\rho\sigma}\eta^{\lambda\kappa}) N_{\alpha \beta \lambda \kappa}(z,w) \notag\\
&\quad =4m^2 E_{0a0b}^{x,\mu\nu} E_{0c0d}^{y,\rho\sigma} \int \frac{d^4p}{(2\pi)^4} e^{ip_\mu (x^\mu-y^\mu)}\Big[D(p)+\frac{4N(p)}{|-(p^0+i\epsilon)^2+\bm{p}^2|^2} \Big] \theta(-p^2-4\mu^2)\mathcal{P}_{\mu\nu\rho\sigma} |_{x^\mu=X^\mu(t), y^\mu=X^\mu(t')}  \notag\\
&\quad \ge 8m^2 E_{0a0b}^{x,\mu\nu} E_{0c0d}^{y,\rho\sigma} \int \frac{d^4p}{(2\pi)^4} e^{ip_\mu (x^\mu-y^\mu)}\sqrt{\frac{4D(p)N(p)}{|-(p^0+i\epsilon)^2+\bm{p}^2|^2}} \theta(-p^2-4\mu^2)\mathcal{P}_{\mu\nu\rho\sigma} |_{x^\mu=X^\mu(t), y^\mu=X^\mu(t')}
\notag\\
&\quad \ge 8 m^2  E_{0a0b}^{x,\mu\nu} E_{0c0d}^{y,\rho\sigma} \int \frac{d^4p}{(2\pi)^4} e^{ip_\mu (x^\mu-y^\mu)}\frac{8\pi G_N }{|-(p^0+i\epsilon)^2+\bm{p}^2|} \theta(-p^2-4\mu^2)\mathcal{P}_{\mu\nu\rho\sigma} |_{x^\mu=X^\mu(t), y^\mu=X^\mu (t')}
\notag \\
&\quad = \frac{64 \pi m^2}{m_p^2}  E_{0a0b}^{x,\mu\nu} E_{0c0d}^{y,\rho\sigma} \int \frac{d^4p}{(2\pi)^4} e^{ip_\mu (x^\mu-y^\mu)}\frac{\theta(-p^2-4\mu^2)}{|p^2|} \mathcal{P}_{\mu\nu\rho\sigma} |_{x^\mu=X^\mu(t), y^\mu=X^\mu (t')}
\notag
\\
&\quad = \Delta^\text{min}_{abcd}
\end{align}
where the  the arithmetic–geometric mean inequality was used in the first inequality, and the deocoherece-diffusion tradeoff \eqref{NDp} was applied in the second inequality. 
We also took the limit $\epsilon \rightarrow 0$. 
Explicitly, $\Delta^\text{min}_{abcd}$ is 
\begin{align}
&\Delta^\text{min}_{abcd}(t,t') \notag\\
&\quad =
 \frac{16m^2 (4\pi G_N)^2}{D_0^{\mathrm{\,Bia}}}\int \frac{d^4p}{(2\pi)^4}\, \frac{ e^{-ip^0(t-t')} \theta((p^0)^2-\bm{p}^2-4\mu^2)}{|-(p^0)^2+\boldsymbol{p}^2|}   \bigg[ (p^0)^4\bigg(\delta_{ac}\delta_{bd} + \delta_{ad}\delta_{bc} -\frac{2}{3}\delta_{ab}\delta_{cd}\bigg) +\frac{4}{3}p_ap_bp_cp_d \notag\\
    &\quad-
    (p^0)^2 \bigg(p_ap_c\delta_{bd} + p_ap_d\delta_{bc} + p_bp_c \delta_{ad} + p_bp_d\delta_{ac} -\frac{2}{3}(p_ap_b\delta_{cd}+p_cp_d\delta_{ab})\bigg)  \bigg].
\end{align} 
For evaluating the power spectral density computed from $\Delta^\text{min}_{abcd}$, the following integrals, 
\begin{align}
    &\int d^3p \, \frac{\theta((p^0)^2-\bm{p}^2-4\mu^2)}{|-(p^0)^2+\bm{p}^2|} \, =4\pi \Big\{-\sqrt{(p^0)^2-4\mu^2}+p^0 \text{Arccoth}\Big[\frac{p^0}{\sqrt{(p^0)^2-4\mu^2}}\Big] \Big\}\theta(p^0-2\mu), \label{intE14}
    \\
    &\int d^3p \, \frac{\theta((p^0)^2-\bm{p}^2-4\mu^2)}{|-(p^0)^2+\bm{p}^2|} \, p_ap_b
    =\frac{4\pi}{3} \Big\{-\frac{4}{3} [(p^0)^2+\mu^2]\sqrt{(p^0)^2-4\mu^2}+(p^0)^3 \text{Arccoth}\Big[\frac{p^0}{\sqrt{(p^0)^2-4\mu^2}}\Big] \Big\}\theta(p^0-2\mu)\, \delta_{ab} , \label{intE15}\\
   &\int d^3p\, \frac{\theta((p^0)^2-\bm{p}^2-4\mu^2)}{|-(p^0)^2+\bm{p}^2|} p_a p_b p_c p_d 
   \notag
   \\
   &\quad =\frac{4\pi}{15}\Big\{-\frac{1}{15}(23(p^0)^4-44(p^0)^2 \mu^2+48\mu^4) \sqrt{(p^0)^2-4\mu^2}+(p^0)^5 \text{Arccoth}\Big[\frac{p^0}{\sqrt{(p^0)^2-4\mu^2}}\Big] \Big\} \theta(p^0-2\mu)\notag
   \\
   &\quad \quad \times  (\delta_{ab}\delta_{cd}+\delta_{ac}\delta_{bd}+\delta_{ad}\delta_{bc}), \label{intE16}
\end{align}
are useful. 
By a straight forward calculation, the minimum power spectral density is given as 
\begin{align}
    (s^h_{x,\text{env}})^2
    &=\frac{1}{m^2 \omega^4} \int dt e^{i\omega t} \Delta^\text{min}_{xxxx} (t,0)
    \notag 
    \\
    &=
    \frac{-2}{27\pi m_p^2}
    \theta(\omega-2\mu)
    \left[
        \frac{\sqrt{\omega^2-4\mu^2}}{5\omega^4}
        \left(
        24\mu^4
        -2\mu^2\omega^2
        +14\omega^4
        \right)
        -3 \omega \,
        \mathrm{Arccoth}\!\left(
        \frac{\omega}{\sqrt{\omega^2-4\mu^2}}
        \right)
    \right].
\end{align}

\bibliography{reference}

\end{document}